\documentclass[twocolumn,aps,pre,footinbib,floatfix,preprintnumbers,amsmath,amssymb,superscriptaddress,10pt,colorlinks]{revtex4-1}

\usepackage{ulem}
\renewcommand{\emph}{\textit}

\usepackage{selinput}
\usepackage{amsmath}
\usepackage{amssymb}
\usepackage{graphicx}
\usepackage{dcolumn} 
\usepackage{bm} 
\usepackage[usenames,dvipsnames]{color}
\usepackage{natbib}
\usepackage{dsfont}
\usepackage{xr}
\usepackage{mathtools}
\usepackage{here}
\usepackage[urlcolor=RoyalBlue,citecolor=RoyalBlue,linkcolor=BrickRed]{hyperref}
\usepackage[mathscr]{euscript}
 
\newcommand{\mean}[1]{\left < #1 \right >}
\newcommand{\abs}[1]{\left | #1 \right |}

\renewcommand{\vec}[1]{\mathbf{ #1 }}

\newcommand{\red}[1]{\textcolor{BrickRed}{#1}}

\begin{document}

\title{\textit{Markovian robots}:~minimal navigation strategies for active particles}

\author{Luis G\'{o}mez Nava}
\affiliation{Universit{\'e} C{\^o}te d'Azur, Laboratoire J.~A.~Dieudonn\'{e},~UMR~7351~CNRS,~Parc Valrose,~F-06108~Nice~Cedex~02,~France}

\author{Robert Gro{\ss}mann}
\affiliation{Universit{\'e} C{\^o}te d'Azur, Laboratoire J.~A.~Dieudonn\'{e},~UMR~7351~CNRS,~Parc Valrose,~F-06108~Nice~Cedex~02,~France}

\author{Fernando Peruani}
\email{peruani@unice.fr}
\affiliation{Universit{\'e} C{\^o}te d'Azur, Laboratoire J.~A.~Dieudonn\'{e},~UMR~7351~CNRS,~Parc Valrose,~F-06108~Nice~Cedex~02,~France}

\begin{abstract}
We explore minimal navigation strategies for active particles in complex, dynamical, external fields, introducing a class of autonomous, self-propelled particles which we call~\textit{Markovian robots}~(MR).  
These machines are equipped with a navigation control system (NCS) that triggers random changes in the direction of self-propulsion of the robots. 
The internal state of the NCS is described by a Boolean variable  that adopts two values. 
The temporal dynamics of this  Boolean variable is dictated by a closed Markov chain~--~ensuring the absence of fixed points in the dynamics~--~with transition rates that may depend exclusively on the instantaneous, local value of the external field. 
Importantly, the NCS does not store past measurements of this value in continuous, internal variables. 
We show that, despite the strong constraints, it is possible to conceive closed  Markov chain motifs that lead to nontrivial motility behaviors of the MR in one, two and three dimensions. 
By analytically reducing the complexity of the NCS dynamics, we obtain an effective description of the long-time motility behavior of the MR that allows us to identify the minimum requirements in the design of NCS motifs and transition rates to perform  complex navigation tasks such as adaptive gradient following, detection of  minima or maxima, or selection of a desired value in a dynamical, external field. 
We put these ideas in practice by assembling a robot that operates by the proposed minimalistic NCS to evaluate the robustness of MR, providing a proof-of-concept that is possible to navigate through  complex information landscapes with such a simple NCS whose internal state can be stored in one bit. 
These ideas may prove useful for the engineering of miniaturized robots.  
\end{abstract}

\date{\today}
\pacs{-}
\maketitle

%%%%%%%%%%%%%%%%%%%%%%%%%%%%%%%%%%%%%%%%%%%%%%%%%%%%%%%%%%%%%%%%%%%%%%%%%%%%%%%%%%%%%%
% main text

\section{Introduction} 
\label{sec:1}

As early as 1959, Feynman discussed the technology transfer from the macro- to the microscale, a highly relevant field of research nowadays in terms of medical applications such as targeted drug delivery and microsurgery~\cite{feynman1960there}.
In recent years, the remarkable advance of nanoscience has made the fabrication of synthetic and molecular machines such as sensors and actuators possible~\cite{bayley2001,hess2017,sanchez2009}. 
Moreover, micrometer-sized devices capable of moving autonomously in a fluid are already a reality. We refer to these microdevices as microrobots. 
Microrobots can transport cargo and invade cells;~healthcare applications for early diagnosis, targeted drug delivery or nanosurgery appear therefore realizable in the not too distant future~\cite{solovev2012,sanchez2016,sanchez2017,Josephe1700362}.  
There exists a large variety of microrobots, rigid and soft ones, whose self-propulsion can be achieved via electrical, chemical or optical stimulation~\cite{solovev2012,sanchez2016,sanchez2017,Josephe1700362}.  
The direction of navigation of these devices can be controlled remotely, for instance via a magnetic field in chemically-driven nanorods~\cite{sanchez2009}.
However, the ultimate goal is to design and fabricate microrobots with a programmable, autonomous navigation system on board integrating sensors, an energy source and actuators. 
At present, the miniaturization of autonomous robots has advanced to the millimeter scale~\cite{churaman2012first,kim_microbiorobotics_2017}. 
Further progress along these lines requires the development of minimal, yet robust algorithms in the sense that they should work reliably in the presence of noise.   

\begin{figure}[t!]
 \begin{center}
  \includegraphics[width=\columnwidth]{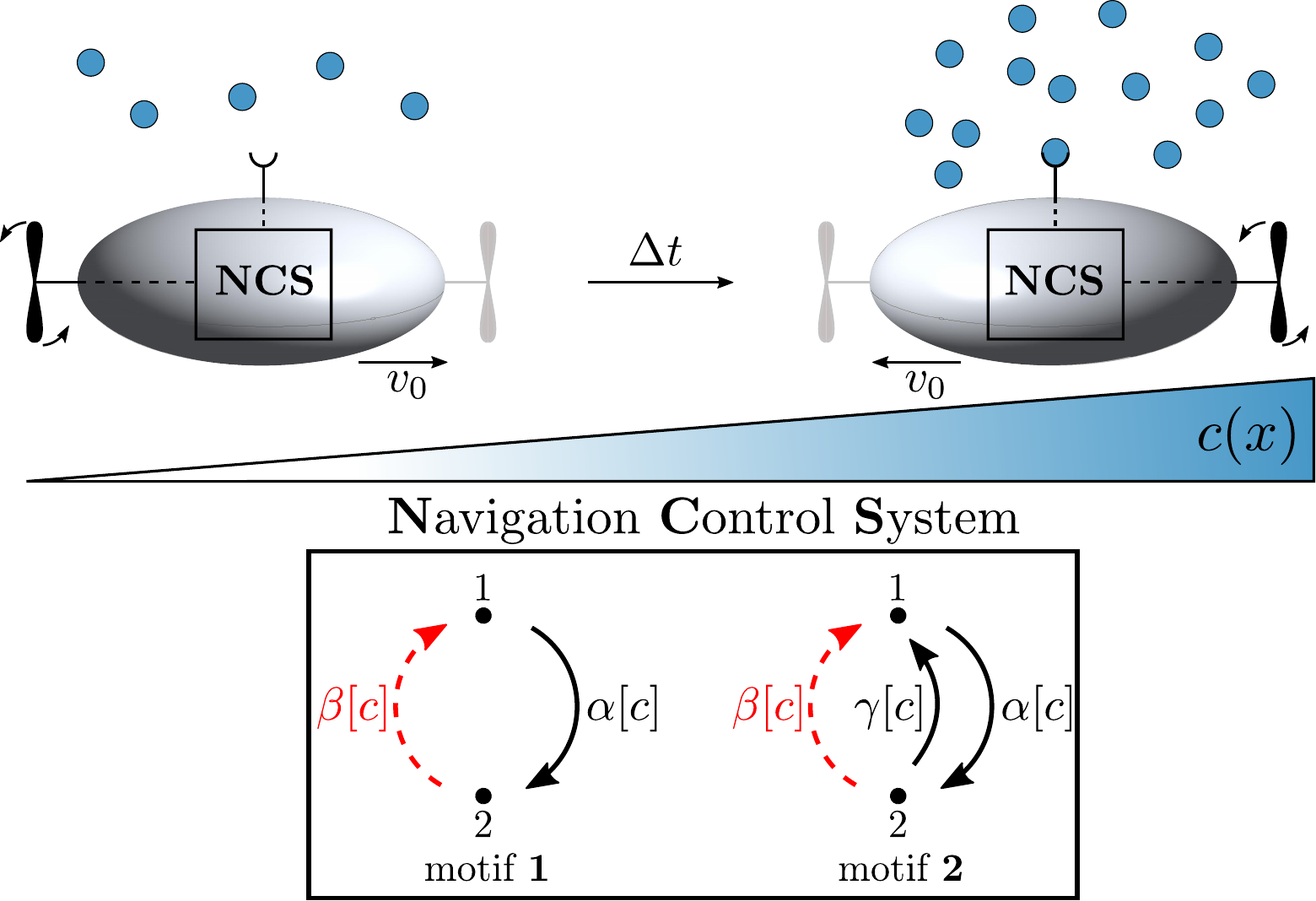}
 \end{center}
 \vspace{-0.35cm} 
 \caption{Illustration of the general dynamics of Markovian robots. A robot that had initially moved from left to right in an external  field~$c(x)$ changed its direction of active motion after some time~$\Delta t$. The navigation control system~(NCS) controls the moving direction of the robot, by triggering reorientation events. It is connected to a sensor which measures the local field values.  The internal state of the NCS is given by a single Boolean variable adopting the values 1 or 2. The NCS dynamics obeys a closed Markov chain, with transition rates that may depend on the value of the external field as measured by the sensor. The red arrow corresponds to the transition that triggers a reorientation event. } 
 \label{fig:scheme}
\end{figure}

Physical properties of small-size objects, e.g. at the microscale, impose technical constraints on the design of microrobots:~viscous forces dominate over inertial ones, fluctuations of thermal origin are not negligible and the instantaneous sensing of external signals can only involve local values, never gradients~\cite{purcell1977life,Berg2008,Tindall2012}.  
For this reason, here we consider a 
class of autonomous, self-propelled particles, which we refer to as~\textit{Markovian robots} (MR), 
that move at constant speed, are subject to fluctuations, and can only sense local values of an external field~(see Fig.\ref{fig:scheme}).  
Notice that we adopt the same constraints small objects are subjected to, but we do not need to assume necessarily that we work at microscopic scales:~the proposed navigation strategies are also of interest for 
macroscopic robots exposed to weakly modulated signals. 
Nevertheless, the long-term motivation of this study is to pave the way for the engineering, in a near future, of tiny, autonomous robots. 
With this intention in mind, we aim at conceiving simple machines that are able to navigate across a complex field~--~providing valuable information clues~--~in an autonomous way with a minimum of information storage capacity.  
Specifically, we equip these machines with a navigation control system (NCS) that triggers random changes in the self-propulsion direction of the robots. 
An essential aspect of the NCS is that it exhibits only two internal states, meaning that the NCS state can be stored in a single  internal Boolean variable  that adopts two values. 
Transitions between these Boolean values are determined by a closed Markov chain, with transition rates that may depend on the instantaneous local value of the external field; see Fig.~\ref{fig:scheme} for sketches of the two  relevant NCS discussed in this work. %, one operating with motif 1 and another one with motif 2. 
Only one of the transition pathways  in the closed two-state Markov chain triggers random reorientation in the moving direction.  
It is worth noticing that the closed-loop nature of the investigated Markov chains ensures the constant resetting of the internal Boolean variable, preventing the presence of fixed points in the dynamics of this variable.  
Importantly, the NCS does not store previous measurements in the form of internal continuous variables, 
preventing a priori any mathematical operation to estimate the gradient of the external field. 
We show that, despite the requested constraints, it is possible to conceive closed  Markov chain motifs that  lead to non-trivial motility behaviors. 
By analytically reducing the complexity in the NCS dynamics, we obtain an effective description of the long-time motility behavior of the MR  
that allows us to  identify the 
minimum requirements in the design of NCS motifs and transition rates to perform  complex navigation tasks such as adaptive gradient following, detection of minima or maxima, or selection of a desired value in a dynamical, external field. 
We show that MR exhibit  non-trivial motility behaviors in one, two and three dimensions.

\begin{figure*}[t!]
 \begin{center}
  \includegraphics[width=\textwidth]{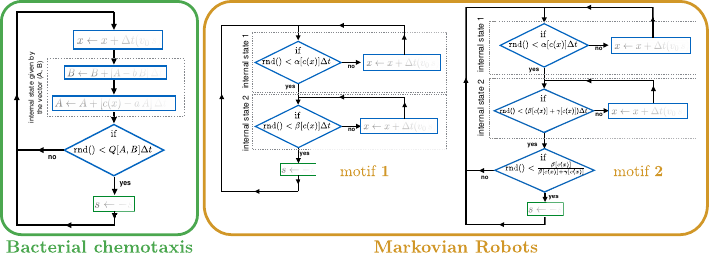}
 \end{center}
 \caption{Flowcharts of the algorithms for bacterial chemotaxis, according to~\cite{Celani2010}, and Markovian robots (MR), motifs 1 and 2. The initial condition is not explicitly shown. Notice the presence of two continuous variables, $A$ and $B$, for bacterial chemotaxis, which are absent for MRs. Further, we point out that the dynamics of $A$ and $B$ evolves towards a fixed point for constant $c(x)$, implying that the internal state in the bacterial chemotaxis model reaches a steady state. In MR, on the other hand, the internal state always oscillates. The symbols~$Q$, $\alpha$, $\beta$, and $\gamma$ refer to transition rates, $\Delta t$ to the time step, $c(x)$ to the value of the external field at position $x$, $v_0$ to the speed, and $s$ to the moving direction ($+1$ or $-1$) in one dimension; $\mbox{rnd}()$ is a uniformly distributed, random number between $0$ and $1$. The definition of $\alpha$, $\beta$, and $\gamma$ are provided in the main text; notice that these rates do only depend on $c(x)$. On the other hand, $Q(A, B)$ is a function of the internal state itself, $Q(A, B)= d_1\,A - d_2\,B$, where $d_1$ and $d_2$ as well as $a$ and $b$ are constant; for details on the bacterial chemotaxis algorithm see~\cite{Celani2010}. } 
 \label{fig:ALGOS}
\end{figure*}

We put these concepts in practice by assembling a macroscopic robot that operates by the proposed NCS 
and is subjected to the constraints indicated above. 
A series of statistical tests allows us to assess the robustness of the proposed minimalistic navigation algorithms. 
The performance of the robot provides solid evidence in favor of the practical interest of these ideas as well as a proof-of-concept that is possible to navigate through a complex information landscape with only 1-bit of memory. 
These ideas may prove of help in the engineering of miniature robots.

The minimalistic navigation strategies discussed here are fundamentally different from bacterial chemotactic strategies~\cite{Berg2008,Tindall2012,Schnitzer1993,Celani2010,Cates2012,Flores2012,chatterjee2011chemotaxis,segall_temporal_1986,deGennes_chemotaxis_2004} as explained in the following.  
In~\cite{Celani2010}, Celani and Vergassola have cleverly shown that bacterial chemotaxis can be described in a Markovian way by enlarging the space of variables, beyond position and velocity variables, to include continuous~(as opposed to Boolean) internal variables. 
The temporal dynamics of these continuous variables obeys a chain of ordinary differential equations, where the first of them depends on the external field.  
The frequency of changes in the moving direction of the bacterium is a function of these variables. 
According to~\cite{Celani2010}, chemotactic behavior is already obtained by keeping the first two of these internal continuous variables.  
The past measurements are encoded by these continuous variables~\cite{Celani2010}, that, from an algorithmic point of view, need to be constantly updated, see Fig.~\ref{fig:ALGOS}. 
These dynamical variables are somehow connected to intracellular chemical species, see~\cite{tu2013} for more details. 
It is worth mentioning that mathematical procedures similar to the ones utilized here have also been used in the context of bacterial chemotaxis, namely a reduction of a complex internal dynamics in order to obtain the effective long-time motility behavior, see e.g.~\cite{Schnitzer1993,Clark9150,Celani2010}. 
However, we stress that the analogy between bacterial chemotaxis and the navigation strategies discussed here is limited to the observation that both strategies are Markovian and make use of internal states.
Notably, there is no direct link between them; see Fig.~\ref{fig:ALGOS} for a comparison at the algorithmic level. 
The differences become evident when looking closely:~here, we discuss navigation strategies that make use of a single internal Boolean variable to describe the internal state of the moving entity, while in~\cite{Celani2010} the internal state of the bacterium is described by~(a minimum of) two continuous variables. 
From this, it is evident that MR have only two possible internal states, while in~\cite{Celani2010} the internal state of a bacterium is given by vector $\mathbf{q}=(A,B)$, with $A\geq0$ and $B\geq 0$ being two internal continuous variables, implying that there is an infinite~(or at least a very large) number of potential internal states.  
Furthermore, the temporal evolution of the continuous variables $A$ and $B$ is given, as indicated above, by a hierarchy of  ordinary differential equations, 
while in MR the temporal evolution of the Boolean variable is given by a  closed Markov chain. 
The differences between both strategies are evident even for a trivial scenario where the external field is constant. The internal variables~$A$ and~$B$ would converge in this scenario to a fixed point and, thus, the internal state reaches a stationary state. 
In contrast, the internal state of MR never converges to a stationary value but oscillates ad infinitum. 
Another indication, how different these strategies are, is the following:~the frequency at which the direction of motion changes in the model~\cite{Celani2010} is a function of the internal state, i.e.~$Q(A,B)$ in Fig.\ref{fig:ALGOS}, while that is not the case for MR. 
In summary, the mathematical structures of both strategies are, analytically and algorithmically, fundamentally different.

While the goal of bacterial chemotaxis models is to understand bacterial navigation, we aim here at conceiving minimal navigation algorithms to engineer simple self-propelled robots. Our intention is to provide new perspectives on the engineering of artificial active particles~\cite{romanczuk_active_2012,vicsek_collective_2012,marchetti_hydrodynamics_2013,menzel_tuned_2015} 
by concentrating on the design of the navigation control system, while  previous studies primarily focused on the design of autonomous self-propulsion mechanisms~\cite{paxton_catalytic_2004,golestanian_designing_2007,kudrolli_swarming_2008,deseigne_collective_2010,bricard_emergence_2013,bricard2015,palacci_living_2013}.

\section{Markovian robots}
\label{sec:2}

In this section, several variants of MR are introduced with a particular focus on their capability of responding to a static, external (scalar) field. 
At first, the general dynamics in space~--~equal for all model variants~--~is formulated. 
Subsequently, several examples of increasing complexity of the internal robot dynamics, which controls the occurrence of reorientation events, are studied.
In particular, an effective Langevin dynamics is derived analytically for each case, which reveals the large-scale robot dynamics in the diffusive limit. 
These concepts are illustrated within a didactic introduction first by means of a simplified version of the model where the reorientation rate is directly a function of the external field. 
% 
%\textcolor{red}{We include the study of the simplest case when the internal variable does not change in time as a didactic introduction to illustrate the series of fundamental concepts used to obtain the long-time dynamics.}

\subsection*{Spatial dynamics}

Throughout, individual robots are assumed to move at constant speed~$v_0$ by means of an active self-propulsion mechanism. 
For simplicity, we focus on one-dimensional systems of linear size~$L$~--~generalizations to higher dimensions are commented on in \red{Section}~\ref{sec:4}. 
In one dimension, the dynamics of the robot is given by 
\begin{align}
	\label{eqn:spat_dyn}
	\dot{x}(t) = v_0 s(t) + \sqrt{2 D_0} \, \xi (t), 
\end{align}
where~$x(t)$ is the position of the robot at time~$t$,~$v_0$ denotes its active speed,~$D_0$ is the bare diffusivity in the absence of active motion~($v_0 = 0$),~$\xi(t)$ abbreviates white Gaussian noise and~$s(t)\in \{ -1 , 1 \}$ indicates the direction of active motion at time~$t$. 
The temporal dynamics of~$s(t)$ is controlled by a \textit{navigation control system}~(abbreviated NCS for short in the following), which we consider to operate 
with one internal Boolean variable that adopts two values. The dynamics of this internal Boolean variable is dictated by 
a closed Markov chain, see Fig.~\ref{fig:scheme} which illustrates several NCS motifs and the robot dynamics. 
%
%The figure includes the very simple situation where the NCS does not even have an internal state (motif 1) that is introduced for didactic purposes. 
%
Notice that the dynamics of the NCS is affected by the external field~$c(x)$ via the $c$-dependency of the transition rates.
In all of these motifs, there is one particular transition leading to state~$1$~(depicted by a red, dashed arrow in Fig.~\ref{fig:scheme}), which triggers a reversal of the driving engine and, thus, induces the inversion of the direction of active motion of the robot:~$s(t) \rightarrow - s(t)$. 
Given a certain NCS motif, we want to understand the motility response of the robot to an external field $c(x)$. This is addressed in the following.

\subsection*{A didactic introduction}

We start by studying the long-time behavior of the simplest possible scenario where the reorientation rate depends directly on the external field, i.e.~there is no internal dynamics. 
% (or equivalently, where the internal states remains always the same). 
%
Let us stress that we use this case as a didactic introduction to illustrate a series of fundamental concepts that will allow us to obtain a simplified long-time dynamics of NCS motifs of higher complexity (NCS motifs 1 and 2).  
Here, reversal events occur at a rate~$\alpha[c]$, which depends on the external signal~$c(x)$. 
The temporal evolution of the system can be expressed in terms of the probabilities~$P^+(x,t)$ and~$P^-(x,t)$ to find a robot at position~$x$ at time~$t$ moving to the right and to the left, respectively.  
The associated Master equation~\cite{gardiner_stochastic_2010} reads:  \hspace{-1cm} 
\begin{subequations}
\label{eqn:master_components_network_1_1} 
\begin{align}
 	\partial_t P^{+} &\!=\! -           v_0 \partial_x P^{+} \!- \alpha[c] P^{+} \!+ \alpha[c] P^{-} \!+ D_0 \partial_x^2 P^{+} \!,  \\
 	\partial_t P^{-} &\!=\! \phantom{-} v_0 \partial_x P^{-} \!- \alpha[c] P^{-} \!+ \alpha[c] P^{+} \!+ D_0 \partial_x^2 P^{-} \!. 
\end{align}
\end{subequations}
By introducing the new variables $P \! \left( x,t \right) \!=\! P^+\! \left( x,t \right) + P^-\! \left( x,t \right)$ and $m\! \left( x,t \right) \!=\! P^+\! \left( x,t \right) - P^-\! \left( x,t \right)$, we recast Eq.~(\ref{eqn:master_components_network_1_1}) into  
\begin{subequations}
\begin{align}
\partial_t P &= -v_0 \partial_x m + D_0 \partial_x^2 P \! , \label{eqn:master_components_network_1_2_a} \\
\partial_t m &= -v_0 \partial_x P + D_0 \partial_x^2 m - 2\alpha[c] m . \label{eqn:master_components_network_1_2_b} 
\end{align}
\end{subequations}
The variable of interest is~$P \! \left( x,t \right)$ representing the probability to find the robot at position~$x$ at time~$t$.  
Since the local dynamics of~$m \! \left( x,t \right)$~[Eq.~(\ref{eqn:master_components_network_1_2_b})] is faster than~$P \! \left( x,t \right)$~[Eq.~(\ref{eqn:master_components_network_1_2_a})] and we are interested in the long-time behavior of the latter, we approximately set~$\partial_t m \approx 0$, enabling us to express~$m \approx - \frac{v_0}{2 \alpha} \partial_x P$ to lowest order in spatial gradients. 
Inserting this expression into Eq.~(\ref{eqn:master_components_network_1_2_a}) yields the following effective equation for the density:  
\begin{equation}
\label{eqn:master_components_network_1_3} 
\begin{aligned}
	\partial_t P & = \partial_x \! \left[ \left ( \! D_0 \!+\! \frac{v_0^2}{2 \alpha[c]} \right ) \! \partial_x P \right] \\ 
 &= - \partial_x\! \left[ P \partial_x\! \left( \frac{v_0^2}{2 \alpha[c]} \right) \right] \! + \partial_x^2 \!\left[ P \left ( \! D_0 \!+\! \frac{v_0^2}{2 \alpha[c]} \right ) \right] \! .
\end{aligned}
\end{equation}
All details related to the reorientation dynamics were coarse-grained by deriving the effective scalar equation~(\ref{eqn:master_components_network_1_3}) for~$P \! \left( x,t \right)$. 
This approach is valid as long as  the mean distance traversed by a robot in between two transitions is shorter than the characteristic scales at which the external field varies.

\subsection*{Effective Langevin dynamics}

Now we consider the inverse problem:~starting with equation~(\ref{eqn:master_components_network_1_3}) for the density~$P \! \left( x,t \right)$, we aim at finding a suitable Langevin equation in Ito's interpretation~\cite{kampen_stochastic_2011,gardiner_stochastic_2010} of the form 
\begin{equation}
 \label{eqn:ansatz_Langevin}
 \dot{x} = f(x) + \! \sqrt{2 D(x)} \, \xi(t) ,
\end{equation}
whose associated Fokker-Planck equation  
\begin{equation} 
\label{eq:FP_generic}
\partial_t P = -\partial_x \Big[ P f(x) \Big] \! + \partial_x^2 \Big [ P D(x) \Big ]  
\end{equation}
for the evolution of the density~$P(x,t)$ is structurally identical to Eq.~(\ref{eqn:master_components_network_1_3})~\footnote{The use of Stratonovich's interpretation leads to a different Fokker-Planck equation~\cite{kampen_stochastic_2011}.}. 
This approach is advantageous in several regards. 
By obtaining an effective drift term~$f(x)$ and an effective diffusion coefficient~$D(x)$, we characterize the transport properties of the MR, encoding the details of the NCS in~$f(x)$ and~$D(x)$. 
The physical interpretation of~$f(x)$ and~$D(x)$ as drift and dispersion, respectively, results from the short-time solution of Eq.~\eqref{eq:FP_generic} for the propagator~\cite{risken_fokker-planck_1996} 
\begin{align*}
	P \! \left( x,t \!+\! \tau| x',t \right) \!\simeq\! \frac{1}{\sqrt{4 \pi D(x') \tau}} \exp \! \left \{ \!- \frac{[x \!-\!x' \!-\! \tau f(x') ]^2}{4 D(x') \tau} \right \} 
\end{align*}
which determines the probability to find a robot at position~$x$ after a \textit{short} observation time~$\tau$ given that it was observed at position~$x'$ at time~$t$. 
Notably, the propagator provides a direct way how to measure the mean local drift or bias~$f(x)$ as well as the position dependent dispersion~$D(x)$.

Knowing drift and diffusion coefficient, we can further determine whether a NCS motif lets the MR display a long-time motility response to the external field as follows. 
The steady state solution~$P_{s}(x)$ of Eq.~(\ref{eq:FP_generic}) for no-flux boundary conditions takes the form 
\begin{equation}
\label{eqn:master_components_network_1_4} 
P_s(x) = \frac{\mathcal{N}}{D(x)} \, \exp \! \left[\int_0^x \! dx' \, \frac{f(x')}{D(x')} \right] 
\end{equation}
with a normalization coefficient~$\mathcal{N}$. 
In general, a MR is said \textit{not} to exhibit a long-term response to a non-constant external field~$c(x)$ if the stationary density \textit{is constant}, i.e., $P_s(x) = P_0 = \mbox{\textit{const.}}$ 
Otherwise, the motif under consideration induces a response in the sense that the coupling to the external field increases or decreases the probability to find a robot in certain areas in space. 
The sign of the derivative of the stationary density distribution is determined by the simple criterion 
\begin{align}
	\partial_x P_s(x) \gtrless 0 \quad \Leftrightarrow \quad f(x) \gtrless \partial_x D(x),   
\end{align}
which follows directly from Eq.~\eqref{eqn:master_components_network_1_4}. 
A constant density~$P_{s}(x)$ requires all $x$-dependencies in Eq.~(\ref{eqn:master_components_network_1_4}) to compensate each other. 
This implies the specific relation~$f(x) = \partial_x D(x)$ between drift and diffusion. 
The nature and form of the response depends on the topology of the motif and the functional form of the rates; this is addressed further below.

Notice that it is possible to obtain a motility response to an external field~$c(x)$ without involving biased motion. 
This is evident from Eq.~(\ref{eqn:master_components_network_1_4}):~if $f(x)=0$ and~$D(x)$ is still a function of~$x$, a nontrivial, stationary density profile will emerge. 
This kind of motility response is known in biology as \textit{chemokinesis}. 
In contrast, directed motion requires a non-vanishing~$f(x)$. In biology, a motility response involving a bias is known as \textit{chemotaxis}.

For the introductory example considered above, the comparison of Eq.~(\ref{eqn:master_components_network_1_3}) and Eq.~(\ref{eq:FP_generic}) reveals 
\begin{align} 
\label{eq:fd_motif1}
	f(x) = \partial_x \! \left[ \frac{v_0^2}{2\alpha[c(x)]} \right] \! , \quad D(x) = D_0 + \frac{v_0^2}{2\alpha[c(x)]},  
\end{align}
which satisfies the above-mentioned relation, i.e.~$f(x) = \partial_x D(x)$, implying~$P_s(x) = P_0 = \mbox{\textit{const}}$. 
We observe that though the diffusion depends on~$x$ and the local drift~$f(x)$ is nonzero and varies over space, there is no long-time motility response~--~the long-time density distribution is flat as noticed when memory kernels were introduced~\cite{Berg2008,Schnitzer1993}. 
Using the terminology of chemotaxis, one can summarize that chemotactic and chemokinetic part compensate each other in this case.

In the following, the powerful approach outlined above is used to express the motility response of MR in the form of Eq.~(\ref{eqn:ansatz_Langevin}) for each motif illustrated in Fig.~\ref{fig:scheme}, where the specific form of~$f(x)$ and~$D(x)$ depends on the motif under consideration.

\subsection*{NCS motif~1: up- \& downgradient motion}

Now, we focus on a more complex scenario where the state of the navigation control system is given by an internal Boolean variable 
that adopts two values: $1$ and $2$. The possible transitions are $1 \to 2$ with rate~$\alpha = \alpha[c]$ and $2 \to 1$ with rate~$\beta = \beta[c]$. 
The latter transition triggers a reversal of the direction of active motion. 
This is motif~$1$ in Fig.~\ref{fig:scheme}. 
Due to the presence of two internal states, we introduce four fields~$P_i^+(x,t)$ and $P_i^-(x,t)$ with $i=\{1,2\}$, which denote the probability to find a robot at position~$x$ at time~$t$ with internal state~$i$ moving to the right~($+$) and to the left~($-$), respectively. 
The temporal evolution of these fields is determined by the following Master equation: 
\begin{subequations}
\label{eqn:master_components_network_2_1} 
\begin{align}
\partial_t P_1^+ \! &= 		-v_0 \partial_x P_1^+ \!- \alpha[c] P_1^+ \!+ \beta[c] P_2^- \! + D_0 \partial_x^2 P_1^+ \! , \\
\partial_t P_1^- \! &= \phantom{-}  v_0 \partial_x P_1^- \!- \alpha[c] P_1^- \!+ \beta[c] P_2^+ \! + D_0 \partial_x^2 P_{1}^-\! , \\
\partial_t P_2^+ \! &= 		-v_0 \partial_x P_2^+ \!- \beta[c] P_2^+ \!+ \alpha[c] P_1^+ \! + D_0 \partial_x^2 P_{2}^+ \! , \\
\partial_t P_2^- \! &= \phantom{-}  v_0 \partial_x P_2^- \!- \beta[c] P_2^- \!+ \alpha[c] P_1^- \! + D_0 \partial_x^2 P_{2}^- \!.
\end{align}
\end{subequations}
By introducing the change of variables $P_i = P_i^+ + P_i^-$ and $m_i = P_i^+ - P_i^-$, we recast Eq.~(\ref{eqn:master_components_network_2_1}) into two groups of equations for the densities~$P_i \! \left( x,t \right)$,  
\begin{subequations}
\label{eqn:master_components_network_2_2} 
\begin{equation}
\begin{aligned}
\label{eq:22_1}
\partial_t P_1 &= -v_0 \partial_x m_1 \!-\! \alpha[c] P_1 \!+\! \beta[c] P_2 \!+\! D_0 \partial_x^2 P_1, \\
\partial_t P_2 &= -v_0 \partial_x m_2 \!+\! \alpha[c] P_1 \!-\! \beta[c] P_2 \!+\! D_0 \partial_x^2 P_2, 
\end{aligned}
\end{equation}
and for the differences~$m_i \! \left( x,t \right)$, 
\begin{equation}
\begin{aligned}
\label{eq:22_3}
\partial_t m_1 &= -v_0 \partial_x P_1 \!-\! \alpha[c] m_1 \!-\! \beta[c] m_2 \!+\! D_0 \partial_x^2 m_1, \\
\partial_t m_2 &= -v_0 \partial_x P_2 \!+\! \alpha[c] m_1 \!-\! \beta[c] m_2 \!+\! D_0 \partial_x^2 m_2 .
\end{aligned}
\end{equation}
\end{subequations}
This seemingly innocent change of variables simplifies the problem substantially. If spatial derivatives in Eqs.~(\ref{eqn:master_components_network_2_2}) were absent, Eqs.~(\ref{eq:22_1}) would decouple completely from Eqs.~(\ref{eq:22_3}). 
Further, we note that the eigenvalues associated to the local dynamics of Eqs.~(\ref{eq:22_1}) are $\lambda_1=0$ and $\lambda_2=-(\alpha+\beta)$, while the real parts of those associated to Eqs.~(\ref{eq:22_3}) are both negative, i.e.~$\Re[\lambda_3],\, \Re[\lambda_4]<0$. 
The lesson is:~in the long-wavelength limit, there is one eigenvector whose temporal evolution is slow while the other three are fast.
Accordingly, we can define a new set of four fields by linear combination of those in Eqs.~(\ref{eqn:master_components_network_2_2}) in such a way that only one of those fields is slow.
Due to number conservation, the total density~$P=P_1+P_2$, which is the primary quantity of interest, is the slow field~($\lambda_1=0$). 
In order to reduce Eqs.~(\ref{eqn:master_components_network_2_2}) to the density dynamics, 
we request local equilibrium and take~$\partial_t m_1 = \partial_t m_2 = 0$, allowing us to  
express all fields as function of $P$ and spatial derivatives of it (see \red{Appendix}~\ref{app:C} for further details).  
By keeping all terms up to second order spatial derivatives of the density~$P$, we obtain an effective Fokker-Planck equation, cf.~Eq.~(\ref{eq:FP_generic}), where~$f(x)$ and~$D(x)$ adopt the form
\begin{subequations}
\label{eqn:master_components_network_2_4}
\begin{align}
\! f(x) &= \frac{v_0^2}{2(\alpha + \beta)} \left[ (\beta - \alpha) \partial_x \! \left( \frac{1}{\alpha} \right) + (\beta + \alpha)\partial_x \! \left( \frac{1}{\beta} \right) \right]\! , \! \! \\
\! D(x) &= D_0 + \frac{v_0^2}{2} \! \cdot \! \frac{\alpha^2 + \beta^2}{\alpha \beta (\alpha + \beta)} . \! \! 
\label{eqn:master_components_network_2_4b}
\end{align}
\end{subequations}
We highlight that~$\alpha[c] = \beta[c]$ yields the relation~$f(x) = \partial_x D(x)$ and, thus, the stationary density~$P_s(x)$ would be a constant according to Eq.~(\ref{eqn:master_components_network_1_4}). 
In other words, we learn that we need to require $\alpha[c] \neq \beta[c]$ and at least one of the rates should depend on $c(x)$ in a nontrivial way in order to design robots that respond to the external field~$c(x)$.
In the spatially homogeneous case, the diffusion coefficient~[Eq.~(\ref{eqn:master_components_network_2_4b})] reduces to the expression, which was derived in Ref.~\cite{grossmann_diffusion_2016}.

\begin{figure}[tb]
	\begin{center}
  		\includegraphics[width=\columnwidth]{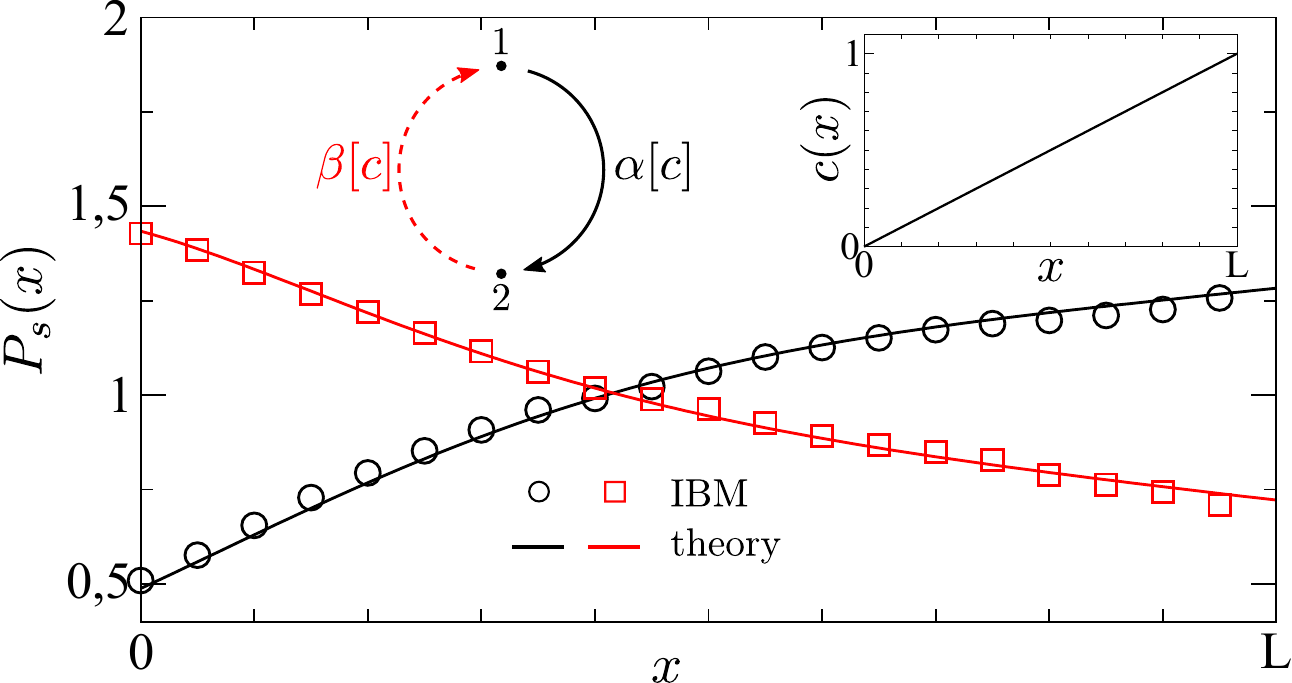}
	\end{center}
	 \vspace{-0.35cm}
	 \caption{The motility response of MR~--~controlled by NCS motif~$1$ as shown, cf.~Fig.~\ref{fig:scheme}~--~to an external field~$c(x) = x/L$~(see inset). The main figure illustrates the stationary probability distributions~$P_{s}(x)$ for two variants of the internal robot dynamics. In the first case~($\alpha[c] \! = \! 9c + 1$, $\beta[c] \!=\! 5$), robots tend to accumulate upgradient in the long-time limit~(circles). In contrast, robots accumulate on the opposite side if the two transitions are interchanged~($\alpha[c] \!=\! 5$, $\beta[c] \! \!= 9c \!+\! 1$) as shown by squares. Points denote individual-based model simulations~(robot number~$N = 10^4$). Lines correspond to the approximative analytical solution~[Eq.~(\ref{eqn:master_components_network_1_4})] where the respective functional forms of~$f(x)$ and~$D(x)$ were inserted~[Eqs.~(\ref{eqn:master_components_network_2_4})]. Further parameters:~$L~=~1$, $v_0 = 0.01$, $D_0 = 0$; reflecting boundary conditions. }
	\label{fig:profile}
\end{figure}

The potentially simplest example that leads to upgradient motion of MR is $\alpha[c] \propto c(x)$ and $\beta[c] = \beta_0$, where~$\beta_0$ is a constant, see Fig. \ref{fig:profile}. 
It is interesting to observe that the robots move downgradient if we make the opposite choice, namely~$\alpha = \mbox{\textit{const.}}$ and~$\beta \propto c(x)$.  
Thus, the previous discussion reveals how the field-dependence of the transition rates controls whether robots tend to move up- or downgradient.

Notably, both types of robots are entirely indistinguishable in spatially homogeneous environments; it is therefore \textit{a priori} impossible to infer information about the type of response on the basis of measurements, which are performed in spatially homogeneous external fields, i.e.~$c(x) = c_0 = \mbox{\textit{const.}}$ at different levels of~$c_0$. 
To be concrete, consider the following gedankenexperiment: two types of robots, robots of type~$A$ with $\alpha[c] \! = \! 9 c + 1$ and $\beta \!=\! 5$ and robots of type $B$ with~$\alpha \!=\! 5 $ and $\beta[c] \! = \!9 c + 1$  exposed to the same external field $c(x)$; this scenario is depicted in~Fig.~\ref{fig:profile}.  
If $c(x)$ corresponds to an homogeneous environment such that $c(x) = c_0$, with $c_0$ being a constant, the diffusion coefficient and the mean rate of reversals are identical for both robot types; notice that the latter increases with the field value~$c_0$. 
One could easily be misled to think that robots tend to accumulate in those regions in space where the reversal rate is high, leading to an effective ``trapping'' of the robots in those regions.  
However, the distinct long-time behaviors of robots of type A and B provide clear evidence against this simplified picture. While robots of type A 
tend to move upgradient and accumulate close to $x=L$, robots of type B tend to move downgradient and accumulate close to $x=0$, see Fig.~\ref{fig:profile}. 
Despite this evident qualitative difference, both types exhibit a higher reversal rate close to $x=L$. 
This finding, i.e. the existence of different motility responses for robots of type A and B, highlights 
how subtle and nontrivial the impact of the NCS design is on the long-time motility response of robots: by exchanging the functional form of the transitions from $1 \to 2$ and $2 \to 1$, we switch from up- to downgradient motion. 

In addition, the previous analysis reveals that robots navigating by NCS motif~$1$ exhibit a motility response that involves both, directed motion and position-dependent diffusion, i.e.~it is neither purely chemotactic nor purely chemokinetic, but involves a combination of both, in the sense that the force~$f(x)$ and the diffusion coefficient~$D(x)$ are non-vanishing functions, which depend on the position via~$c(x)$.

\subsection*{NCS motif~2: adaptation}

\begin{figure*}[htb]
 \begin{center}
  \includegraphics[width=\textwidth]{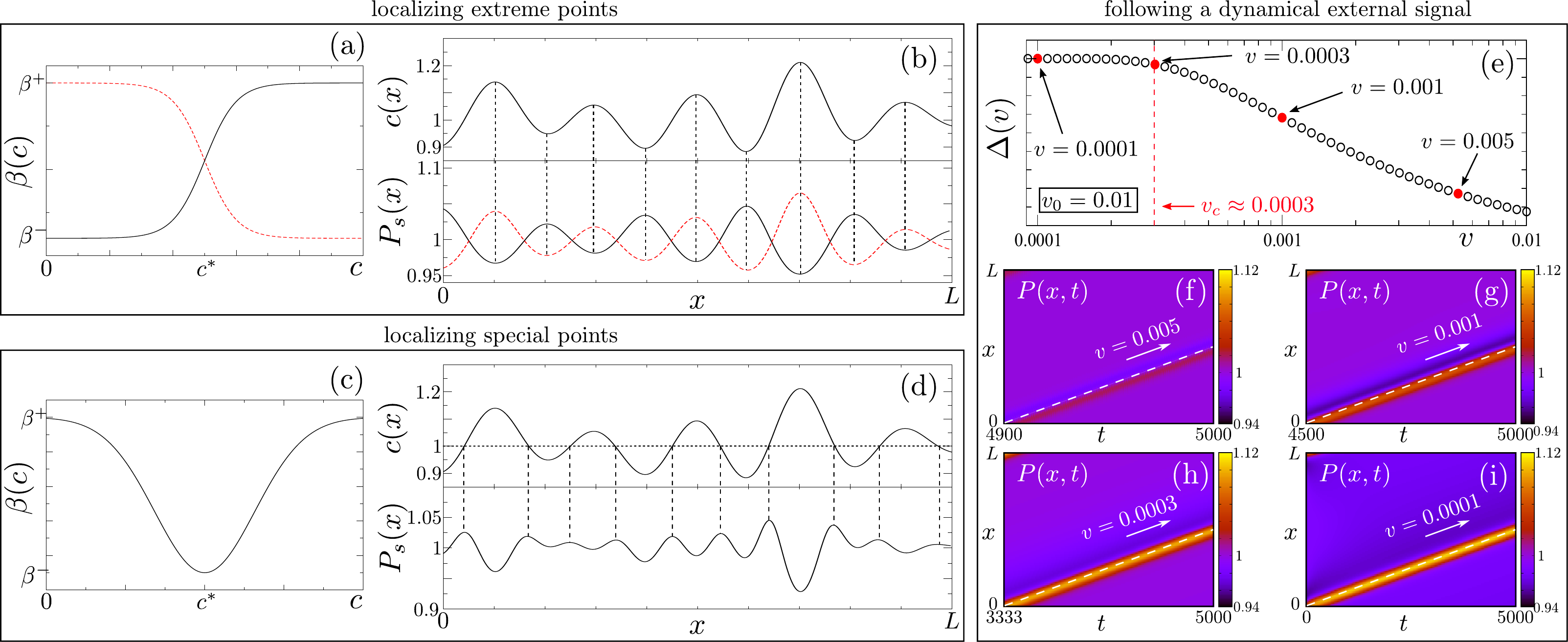}
 \end{center}
 \vspace{-0.35cm}
 \caption{Illustration of complex tasks performed by suitably tuned robots controlled by the adaptive NCS motif~$2$. The detection of maxima and minima in a complex landscape $c(x)$ is shown at the bottom of panel~(b) [$c(x)$ is displayed at the top of panel (b)]. The corresponding functional dependencies of~$\beta[c]$ are shown in~(a):~for increasing~$\beta[c]$, minima are detected~(black, solid curves) whereas robots accumulate around maxima for decreasing~$\beta[c]$~(red, dashed lines). Moreover, the accumulation of robots around a preferred external field value~[dotted line in~(d)] is demonstrated for the functional dependence~$\beta[c]$ shown in~(c). As a third example, robots chasing for a moving signal~(white, dashed line) are depicted in panels~(e)-(i). Space-time plots~(f)-(i) reveal that robots become less responsive to a moving signal above a critical speed~$v_c$, estimated by~Eq.~(\ref{eqn:crit_v_est}), which is shown by a vertical red~(dashed) line in (e). Further, the performance is quantified in~(e) where~$\Delta$ denotes the deviation of the robot density from the spatially homogeneous distribution. Parameters:~$c^* = 1$, $w=0.5$, $\beta^- = 1$, $\beta^+ = 10$, $v_0=0.01$, $L=1$, $D_0 = 0$; see main text for the functional forms of the rate~$\beta[c]$. Boundary conditions:~reflecting in~(b) and~(d), periodic in~(e)-(i). } 
 \label{fig:tasks}
\end{figure*}

We introduce NCS motif~$2$~(cf.~Fig.~\ref{fig:scheme}) to obtain robots whose motility response is purely chemotactic, with a bias resulting from $f(x)$ only. 
Adopting the terminology of bacterial chemotaxis, we call robots whose diffusion coefficient possesses an explicit dependency on $c(x)$, and thus on~$x$ \textit{nonadaptive}, while those with a constant diffusion coefficient are referred to as \textit{adaptive} robots. 
Following this nomenclature, we seek to create chemotactic, adaptive robots. 
Adaptive MR are characterized by the independence of their motility pattern from the intensity of external stimuli in spatially homogeneous environments~--~the diffusion coefficient, for example, is independent of the basis level of the external field. 
We insist that only NCS motif~$2$ can yield adaptive robots; motif~$1$ leads always to nonadaptive motility responses.

Motif~$2$ differs from motif~$1$ by the existence of a backward transition from state~$2$ to~$1$, which does not activate a reversal, and  whose associated transition rate is~$\gamma[c]$. 
Following the analytical procedure outlined before, we first write the equations for~$P_i^{\pm}$ and perform the change of variables~$P_i = P_i^+ + P_i^-$ and~$m_i = P_i^+ - P_i^-$ for all~$i$. 
Again, the dynamics of the~$m_i$'s is fast and, furthermore, the system of equations for the~$P_i$'s contains one fast and one slow mode allowing us, eventually, to reduce the four-dimensional system to the slow dynamics of the density~$P=P_1 + P_2$. 
Keeping derivatives up to second order, we obtain a Fokker-Planck equation of the form given by Eq.~(\ref{eq:FP_generic}) where 
\begin{subequations}
\label{eqn:master_components_network_3_3} 
\begin{align}
\label{eq:3_3_a}
f(x) &= \frac{v_0^2}{2} \left[\partial_x \! \left( \frac{\alpha + \gamma}{\alpha \beta} \right) \! + \frac{(\beta + \gamma - \alpha)}{(\alpha + \beta + \gamma)} \partial_x \! \left( \frac{1}{\alpha} \right) \right] \! ,\\
\label{eq:3_3_b}
D(x) &= D_0 + \frac{v_0^2}{2} \! \cdot \! \frac{\alpha^2 + (\beta + \gamma)^2 + 2\alpha \gamma}{\alpha \beta (\alpha + \beta + \gamma)}.
\end{align}
\end{subequations}
In the limit~$\gamma \to 0$, Eqs.~(\ref{eqn:master_components_network_2_4}) are recovered. 
Again, if all rates are equal, $\alpha[c] \!=\! \beta[c] \!=\! \gamma[c]$, chemotactic and chemokinetic part are related by $f(x) = \partial_x D(x)$ and, thus, the stationary density~$P_s(x)$ is constant.

So far, no restrictions have been imposed on the rates~$\alpha$, $\beta$, and $\gamma$. 
Consequently, the terms~$f(x)$ and~$D(x)$ given by Eqs.~(\ref{eqn:master_components_network_3_3}) are generic for motif~$2$.  
In order to obtain adaptive robots, we want to choose these rates in such a way that~$D(x)$ becomes independent of~$c(x)$, while~$f(x)$ still depends on it. 
With this idea in mind, we define
\begin{subequations}
	\label{eqn:choice:adapt}
\begin{align}
	\alpha[c] &= \frac{\beta_+\beta_-}{\beta[c]}, \\
	\gamma[c] &= \beta_{+} + \beta_{-} - \big (\alpha[c]+\beta[c] \big ). 
\end{align}
\end{subequations}
In order to ensure all rates to be positive, we further choose~$\beta_{-}<\beta[c]<\beta_{+}$, where $\beta_{-}$ and $\beta_{+}$ are positive constants.
By inserting these rates into Eq.~(\ref{eqn:master_components_network_3_3}), we find \hspace{-1cm}
\begin{subequations}
\label{eqn:master_components_network_3_5} 
\begin{align}
 f(x) &= -\frac{v_0^2}{\beta_+ + \beta_-} \partial_x \Big ( \! \ln \beta[c] \Big ) \! = \mu[c] \partial_x c(x), \label{eqn:master_components_network_3_5a} \!\! 
 \\
 D(x) &= D_0 + \frac{v_0^2}{2} \! \cdot \! \frac{\beta_+^2 + \beta_-^2}{\beta_+ \beta_- \left(\beta_+ + \beta_- \right)}.  \label{eqn:master_components_network_3_5b} 
\end{align}
\end{subequations}
Notably, Eq.~(\ref{eqn:master_components_network_3_5b}) is structurally identical to~Eq.~(\ref{eqn:master_components_network_2_4b}) for motif~$1$, however, by definition it is independent of~$c(x)$. 
Further, we defined the response function 
\begin{align}
	\label{eqn:mobility} 
	\mu[c] = -\frac{v_0^2}{\beta_+ + \beta_-}\! \cdot \! \frac{\partial_c \beta[c]}{\beta[c]}. 
\end{align}
Notice  that any function restricting the values of~$\beta[c]$ between~$\beta_-$ and~$\beta_+$ serves our purpose. 
This freedom of choice may be used to design~$\mu \! \left[ c \right]$ according to the desired response.

With the above choice of rates, we obtain purely adaptive, chemotactic robots whose directed motion is controlled by~$f(x)$ only~[cf.~Eq.~(\ref{eqn:master_components_network_3_5})]. 
In the absence of a external field gradient,~$\partial_x c(x)=0$, robots diffuse with a constant diffusion coefficient given by~Eq.~(\ref{eqn:master_components_network_3_5b}) that is independent of the external field value. 
We notice that requesting~$D = \mbox{\textit{const}}.$ is equivalent to fixing the average and variance of the run-time distribution of the robots;~accordingly, their behavior in homogeneous environments of different (constant) field values is microscopically indistinguishable. %  at different external concentrations.  

\section{Performing complex tasks}
\label{sec:3}

In the following, we discuss the possibility of designing MR to perform multiple complex tasks by playing with the response function~$\mu[c]$, cf.~Eq.~(\ref{eqn:mobility}), on the basis of NCS motif~$2$. 
If we define~$\beta[c]$ such that~$\mu[c]>0$ in the interval of interest of field values, robots  move upgradient. 
As a consequence, they accumulate around the maxima of~$c(x)$ in a complex landscape as shown in Figs.~\ref{fig:tasks}\red{(a,b)}. 
This requires~$\beta[c]$ to be a decreasing function of~$c$. 
In addition, we have to make sure that~$\beta[c]$ is bounded by~$\beta \! \in \! (\beta_{-}, \beta_{+})$. 
As an example, we consider~$\beta[c] \!=\! A + B \tanh\!\left[(c - c^*)/w \right]$ where~$A$ and~$B$ are chosen such that~$\beta[c\to 0] = \beta^+$ and~$\beta[{c\to \infty}] = \beta^-$, and where~$w$ and~$c^*$ are constants. 
On the other hand:~if robots are supposed to move downgradient to accumulate in the minima of~$c(x)$, the response function has to be a decreasing function of the signal, $\mu[c]<0$, and, thus,~$\beta[c]$ should be an increasing function of~$c$. 
This can be achieved by using the same functional form as before, but requesting~$\beta[c\to 0] = \beta^-$ and~$\beta[{c\to \infty}] = \beta^+$, cf.~Figs.~\ref{fig:tasks}\red{(a,b)}.

We can further design robots to accumulate at a given value~$c^*$ of the external field as shown in Figs.~\ref{fig:tasks}\red{(c,d)}. 
For this task, we need~$\mu[c]$ to be positive for~$c<c^*$ and negative for~$c>c^*$. 
Fig.~\ref{fig:tasks}\red{(c)} illustrates this type of robot design:~$\beta[c] = A - B \mathscr{N}(c;c^*\!, w^2)$ where~$\mathscr{N}(c;c^*\!, w^2)$ is a Gaussian distribution centered at~$c^*$ and of variance~$w^2$. The coefficients~$A$ and~$B$ are chosen such that~$\beta[c \to c^*] = \beta_{-}$ and~$\beta[c\to0]=\beta_{+}$.

As a final example, we study how robots chase a signal that moves at speed $v$ as shown in Figs.~\ref{fig:tasks}\red{(e)-(i)}. 
The analyzed scenario is analogous to recent bacterial chemotactic experiments 
performed with a moving chemoattractant signal~\cite{Li_barrier_2017}. 
For this purpose, the MR design for the detection of maxima is used, cf.~the discussion of Figs.~\ref{fig:tasks}\red{(a,b)}. 
As a signal, we use a Gaussian distribution that moves at a constant speed $v$ (remember that robots move at constant speed $v_0$).    
There is a critical signal speed~$v_{c}$ above which the robots become decreasingly responsive.
In the limit of high signal speeds~($v \! \gg \! v_c$), robots just diffuse around as they would do in an homogeneous field. 
We quantify the efficiency of robots by the deviation from the homogeneous distribution~$\Delta \propto \int_{0}^L dx \left | P_s(x) - P_0 \right |$, see Fig.~\ref{fig:tasks}\red{(e)}, which is large if robots follow the moving signal and decreases to zero for non-responsive MR. 
Considering a simplified scenario, we derive a rough estimate of the crossover speed beyond which robots cannot follow the moving signal. 
The estimation is based on the idea of a quasi-stationary situation:~the density distribution of robots should have relaxed into the stationary state before the signal has moved to ensure that robots can follow a dynamic signal. 
Imagine first a static field~$c(x) \! = \! \bar{c} \exp[-(x-x_0)^2/(2h^2)]$, where~$x_0$ and~$h$ are constants such that $f(x) \! = \! - \mu[c] c(x) (x-x_0)/h^2 = - \kappa (x - x_0)$; notice that $h$ is a length-scale that characterizes the spatial extent of the gradient.
Assuming that we can linearize around~$x_0$, we approximate~$\kappa \approx \mu [\bar{c}] \bar{c}/h^2$, yielding a linear restoring force~$f(x)$ as if it was coming from a harmonic potential. 
The characteristic relaxation time for an harmonic potential is~$\kappa^{-1}$. 
The critical speed is now given by the product of the relaxation rate and a characteristic size of the signal;~therefore, we estimate that robots can follow any signal that travels at a speed less or equal to 
\begin{align}
	\label{eqn:crit_v_est}
	v_c = h \times \kappa \simeq \frac{\mu \! \left[ \bar{c} \right] \! \bar{c}}{h}. 
\end{align}
For the parameters used in the simulations shown in Fig.~\ref{fig:tasks}, the critical speed yields~$v_c \approx 3 \! \times \! 10^{-4}$, indicated by a red~(dashed) line in Fig.~\ref{fig:tasks}\red{(e)}.

\section{Two \& three-dimensional systems} 
\label{sec:4}

The results obtained so far regarding the motility response of MR, based on a one-dimensional approach, hold true qualitatively in higher dimensions. 
Below, we briefly outline how biased motion of MR can be addressed in higher spatial dimensions within the same theoretical framework and provide a proof of principle. 
Technical details as well as a full account of the general dynamics in two as well as in three dimensions can be found in \red{Appendix}~\ref{app:D} and \red{Appendix}~\ref{app:E}, respectively.

\paragraph*{Two-dimensional case}

\begin{figure*}[tb]
	\begin{center}
		\includegraphics[width=\textwidth]{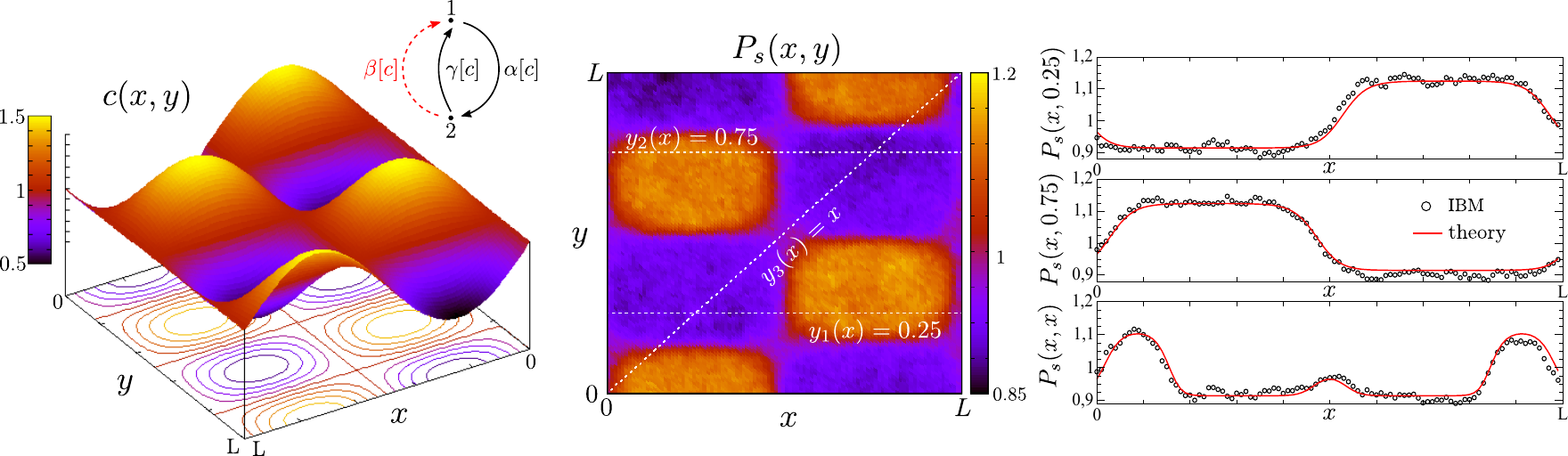}
		\vspace{-0.35cm}
		\caption{Illustration of adaptive MR in two dimensions. In the left panel, the external field~$c(\vec{r}) = 1 + 0.5 \sin \! \left( 2 \pi x \right) \! \cos \! \left( 3 \pi y \right)$ is shown. The middle panel represents the stationary probability density~$P_s \! \left( \vec{r} \right)$ obtained from individual based model~(IBM) simulations. On the right, several cross sections as indicated in the middle panel by white~(dashed) lines are shown in comparison to predictions of the drift-diffusion approximation~[Eq.~\eqref{eqn:dd2d}]. Model specification:~upon reorientation, a robot chooses a new direction of motion from the uniform probability distribution~$g \! \left( \varphi \right) = 1/(2\pi)$, such that~$\mathcal{G} = 0$; adaptive NCS motif~$2$, cf.~Eqs.~\eqref{eqn:choice:adapt}; $\beta[c]$ as shown in Fig.~\ref{fig:tasks} (red line) and the corresponding comments in the main text~($\beta_{-} = 1$, $\beta_{+} = 10$). Other parameters:~$L=1$, $v_0 = 0.01$, $D_0 = 0$, $D_r = 0$, $N = 10^4$ robots in IBM simulations; reflecting boundary conditions. }
		\label{fig:2d}
	\end{center}
\end{figure*}

The equation of motion of a MR in two dimensions reads
\begin{equation}
	\dot{\mathbf{r}}(t) = v_0 \hat{\mathbf{s}}[\varphi(t)] + \!\!\;\sqrt{2 D_0} \, \boldsymbol{\xi} \! \left( t \right) \! ,
\end{equation}
where $\mathbf{r}(t)$ is the position of a robot in two dimensions and $\hat{\mathbf{s}}[\varphi] = (\cos \varphi, \sin \varphi)$ is a unit vector pointing in the direction of motion parametrized by the polar angle~$\varphi$.  
The polar angle may undergo a stochastic, rotational dynamics due to small-scale spatial heterogeneities, thermal fluctuations or temporal variations of the active driving force~\cite{mikhailov_self_1997,peruani_self_2007,romanczuk_brownian_2011,chepizhko2013}: 
\begin{align}
	\dot{\varphi} (t) = \sqrt{2D_r} \, \eta(t). 
\end{align}
The noise amplitude~$D_r$ parametrizes the persistence of trajectories during run phases, and~$\eta(t)$ denotes white, Gaussian noise.

The internal robot dynamics, which controls abrupt changes of the direction of motion, is determined as before by a certain NCS. 
The reorientation of robots may be implemented in several ways:~the new direction of active motion could be selected from a probability distribution of reorientation angles, representing, for example, a cone centered around the previous orientation or it could be chosen uncorrelated with respect to the previous direction of motion. The qualitative behavior is independent of this choice.

At first, we put forward a heuristic argument valid for low angular noise intensities to illustrate why the results derived so far based on one-dimensional systems should hold in higher dimensions. 
Consider a robot equipped with some NCS, which controls the moments in time when the robot selects a new direction of motion from a certain probability distribution. 
The velocity of a robot may always be divided into its components parallel and perpendicular to the gradient orientation. 
The upgradient climbing speed~$v_{\perp}$ is a random variable, which changes at each reorientation event.
Thus, the speed has to be rescaled to obtain an average climbing speed. 
Furthermore, not every reorientation inevitably leads to a reversal of the direction of active motion with respect to the gradient orientation. 
Upon reorientation, the projection of the new direction of active motion onto the gradient is positive or negative with equal probability~($p=1/2$) if the direction of self-propulsion of a robot at each reorientation event is chosen randomly from the interval~$[-\pi,\pi)$ in two dimensions;~in general, there is a reversal probability~$p_r$ that the robot moves upgradient~(downgradient) given that it was moving downgradient~(upgradient) before the reorientation. 
These arguments indicate that it is always possible to come up with an effective one-dimensional description~--~in the sense of a projection~--~for the motion along the local gradient orientation, which is analogous to the problem considered in previous sections.

We now turn to a quantitative analysis of the problem in two dimensions. 
For the sake of concreteness, we formulate the problem for adaptive robots as discussed in \red{Section}~\ref{sec:2}, which are controlled by NCS motif~$2$, cf.~Fig.~\ref{fig:scheme}. 
In contrast to one dimension, where only two directions of motion~(denoted by~$\pm$) are possible, a continuum of orientations parametrized by the polar angle~$\varphi$ exists in two dimensions. 
Therefore, the probability densities~$P_i^{(\pm)} \! \left( x,t \right)$ are replaced by the probability densities~$\mathcal{P}_i \! \left( \vec{r},\varphi,t \right)$ to find a robot at position~$\vec{r}$ in state~$i$, moving into direction~$\varphi$ at time~$t$. 
We introduced the new symbol~$\mathcal{P}_i \! \left( \vec{r},\varphi,t \right)$ in order to avoid confusion with the probability density
\begin{align}
	\label{eqn:dens:2d:gen}
	P_i \! \left( \vec{r},t \right) = \int_{\! - \pi}^{\pi} d \varphi \, \mathcal{P}_i \! \left( \vec{r},\varphi,t \right) 
\end{align}
to find a robot at a certain position~$\vec{r}$ at time~$t$ in state~$i$, independent of its direction of motion. 
Altogether, the full set of Master equations for robots controlled by NCS motif~$2$ in two dimensions reads
\begin{widetext}
\begin{subequations}
\label{eqn:master:2d}
  \begin{align}
	  \partial_t & \mathcal{P}_1 \! \left( \vec{r},\varphi,t \right)   =  - v_0 \!\:\hat{\mathbf{s}}[\varphi]  \cdot  \nabla \mathcal{P}_1 + D_r \partial_{\varphi}^2 \mathcal{P}_1 + D_{0} \Delta \mathcal{P}_1- \alpha[c] \mathcal{P}_1 + \gamma[c] \mathcal{P}_2 + \beta[c] \int_{\!-\pi}^{\pi} \!\! d \varphi' \, g \! \left( \varphi - \varphi' \right) \mathcal{P}_2 \! \left( \vec{r},\varphi' \! ,t \right) \! , \\
	\partial_t & \mathcal{P}_2\! \left( \vec{r},\varphi,t \right)   =  - v_0 \!\:\hat{\mathbf{s}}[\varphi]  \cdot  \nabla \mathcal{P}_2 + D_r \partial_{\varphi}^2 \mathcal{P}_2 + D_{0} \Delta \mathcal{P}_2 - \Big ( \beta[c] + \gamma[c] \Big ) \mathcal{P}_2 + \alpha[c] \mathcal{P}_1 . 
  \end{align}
\end{subequations}
\end{widetext}
The details of the stochastic reorientation event, triggered by the NCS, are determined by the probability density function~$g(\varphi)$. For unbiased reorientations, this function should be symmetric:~$g(-\varphi) = g(\varphi)$.

Just as in one dimension, the total density dynamics is a slow quantity since it is locally conserved. It is therefore possible to reduce the set of Master equations to the Fokker-Planck equation
\begin{align}
\label{eqn:FP:2d:t}
	\partial_t P \! \left( \vec{r},t \right) = - \nabla \! \cdot \! \Big[ \vec{f} \! \left( \vec{r} \right) \! P \! \left( \vec{r},t \right) \! \Big] \! + \Delta \Big[ D \! \left( \vec{r} \right) \! P \! \left( \vec{r},t \right) \! \Big] 
\end{align}
for the density~$P \! \left( \vec{r},t \right) = \sum_i P_i \! \left( \vec{r},t \right)$. 
Technically, the derivation follows the same logic as in one dimension. At first, the dynamics of the probability densities~$P_i \! \left( \vec{r},t \right)$, cf.~Eq.~\eqref{eqn:dens:2d:gen}, are derived by integration of Eqs.~\eqref{eqn:master:2d} over the angular variable~$\varphi$. 
Those equations are coupled to the flux, which is determined by the local order parameter
\begin{align}
	\mathbf{m}_i \! \left( \vec{r},t \right) \! = \!\! \int \!\! d \varphi \,  \hat{\mathbf{s}} \! \left[ \varphi \right] \! \mathcal{P}_i \! \left( \vec{r},\varphi,t \right)
	\!=\! \!\int \!\!d \varphi 
	\begin{pmatrix}
		\cos \varphi \\ \sin \varphi
	\end{pmatrix} \! 
	\mathcal{P}_i \! \left( \vec{r},\varphi,t \right) 
\end{align}
in two dimensions. The fields~$\mathbf{m}_i$, which replace $m_i$ from the one-dimensional discussion, may again be adiabatically eliminated to obtain a reduced set of equations for the densities~$P_i \! \left( \vec{r},t \right)$. Finally, the density dynamics follows by assuming local equilibrium. Details on this derivation are summarized in \red{Appendix}~\ref{app:D}. 
For the example considered above, namely adaptive MR with NCS motif~$2$, one obtains the local drift
\begin{align}
	\label{eqn:drift:2d}
	\vec{f} \! \left( \vec{r} \right) &= - \Lambda_2 \nabla \Big( \! \ln \beta [c] \Big)
\end{align}
with the parameter-dependent prefactor
\begin{align}
	\label{eqn:lambda_coeff}
	\!\! \Lambda_2 \!=\! \frac{v_0^2}{2} \! \cdot \! \frac{\beta_{+}\beta_{-} \! \left( 1 - \mathcal{G} \right)}{ \left ( \beta_{+} \! + \! \beta_{-} \right ) \! \cdot \! \Big [ \! \left( \beta_{+} \! + \! D_r \right) \left( \beta_{-} \! + \! D_r \right) \!-\! \beta_{+} \beta_{-} \mathcal{G} \, \Big ]} . \!\!
\end{align}
Further, the constant diffusion coefficient reads 
\begin{align}
 	\label{eqn:diff:2d}
	\!D \!=\! D_0 \!+\! \frac{v_0^2}{2} \! \cdot \! \frac{\beta_{+}^2 \!\!\;+\!\!\; \beta_{-}^2 \!\!\;+\!\!\; \beta_{+}\beta_{-}\!\left( 1 \!\!\;+\!\!\; \mathcal{G} \right) \!\!\;+\!\!\; D_r \! \left( \beta_{+} \!\!\;+\!\!\; \beta_{-} \right)}{ \left ( \beta_{+} \! + \! \beta_{-} \right ) \! \cdot \! \Big [ \! \left( \beta_{+} \! + \! D_r \right) \left( \beta_{-} \! + \! D_r \right) \!-\! \beta_{+} \beta_{-} \mathcal{G} \, \Big ] } .
\end{align}
In these effective transport quantities, the mean cosine of the reorientation distribution~$g(\varphi)$ was abbreviated by~$\mathcal{G}$: 
\begin{align}
	\mathcal{G} = \mean{\cos \varphi} = \int_{-\pi}^{\pi} d\varphi \, g \! \left( \varphi \right) \cos \varphi \,  \! .
\end{align}
These results constitute a straightforward generalization of the results for the one-dimensional case.

The stationary density distribution, i.e.~the stationary solution of the Fokker-Planck equation~\eqref{eqn:FP:2d:t}, for adaptive MR in two dimensions reads 
\begin{align}
	\label{eqn:dd2d}
	P_s \! \left( \vec{r} \right) = \frac{\!\mathcal{N}}{\big (\beta \! \left[ c \right] \hspace{-0.02cm}\big )^{ \hspace{-0.05cm} \Lambda_2 / D}} . 
\end{align}
It is illustrated in Fig.~\ref{fig:2d} together with a comparison to individual based model simulations.

\paragraph*{Three-dimensional case}

For the sake of completeness, we finally consider MR in three spatial dimensions. 
Their transport characteristics turn out to be marginally different from those in two dimensions, as explained below. 
This implies in particular that all qualitative statements made above hold true in three dimensions as well. 
However, there are some technical complications as a consequence of the three dimensional motion regarding the implementation of angular fluctuations as well as the angular reorientation, which are explained below for this reason.
Again, adaptive MR controlled by NCS~motif~$2$ are considered for simplicity as an example. All details concerning the general dynamics of MR in three dimensions can be found in \red{Appendix}~\ref{app:E}.

The dynamics in space for MR in three dimensions, 
\begin{equation}
	\dot{\mathbf{r}}(t) = v_0 \hat{\mathbf{s}} + \! \sqrt{2 D_0} \, \boldsymbol{\xi} \! \left( t \right) \! ,
\end{equation}
is unchanged with respect to previous cases. 
However, the orientation of the active driving force, determined by the unit vector~$\hat{\vec{s}}$, has to be parametrized differently in three dimensions.
One could, for example, use spherical coordinates~$\hat{\vec{s}} = \left( \sin \theta \cos \varphi, \sin \theta \sin \varphi, \cos \theta \right)$, where~$\theta \in [0,\pi]$ and $\varphi \in \left [-\pi,\pi \right)$.  
The angular dynamics for unbiased, orientational fluctuations reads then as follows:   
\begin{subequations}
	\label{eq:ang_dyn3d}
	\begin{align}
		\dot{\theta}  &= D_r \!\!\: \cot \theta + \! \sqrt{2 D_r} \, \eta_{\theta}(t), \\ 
		\dot{\varphi} &= \sqrt{2 D_r} \:\! \csc\theta \, \eta_{\varphi}(t). 
	\end{align}
\end{subequations}
For a derivation of these equations, see~\cite{grossmann_anistropic_2015}, and for a detailed discussion on Brownian motion in 3D we refer the 
reader to 
\cite{Perrin,yosida1949,doi:10.1143/JPSJ.22.219,Brillinger1997,PhysRevLett.103.068102}.  
All interpretations of multiplicative noise terms in the angular dynamics~[Eq.~\eqref{eq:ang_dyn3d}] are equivalent in this particular case.
From a technical point of view, it is, however, more convenient to use Cartesian coordinates for the director~$\hat{\vec{s}} = \left( s_x,s_y,s_z \right)$, at least for analytical calculations.

Besides the continuous fluctuations of the direction of motion due to rotational diffusion~[Eq.~\eqref{eq:ang_dyn3d}], robots change the orientation of the active driving force in a discontinuous fashion each time that the NCS triggers one of this events:~$\hat{\vec{s}}' \to \hat{\vec{s}}$. 
Given that the previous direction of motion was~$\hat{\vec{s}}'$, a novel orientation~$\hat{\vec{s}}$ is chosen from a transition probability density~$g \! \left(  \hat{\vec{s}} | \hat{\vec{s}}' \right)$. 
Since reorientations are supposed to occur in an unbiased manner, the transition probability density~$g \! \left(  \hat{\vec{s}} | \hat{\vec{s}}' \right)$ can only depend on the scalar product~$\hat{\vec{s}} \cdot \hat{\vec{s}}'$.
Further, the normalization of the director, $\abs{\hat{\vec{s}}} = 1$, has to be preserved. One may, therefore, parametrize
\begin{align}
	g \! \left(  \hat{\vec{s}} | \hat{\vec{s}}' \right) = \frac{\delta \! \left( 1 - \abs{\hat{\vec{s}}} \right)}{2 \pi} \,  \mathscr{H} \! \left( \hat{\vec{s}} \! \cdot \! \hat{\vec{s}}' \right) \! , 
\end{align}
where~$ \mathscr{H}\! \left(  \hat{\vec{s}} \cdot \hat{\vec{s}}' \right)$ is the probability distribution function for the scalar product of the orientations just right before and after a reorientation event.
Put differently, it denotes the probability density for the cosine of the angle~$\psi$ between the vectors~$\hat{\vec{s}}$ and $\hat{\vec{s}}'$, i.e.~$\cos \psi =\hat{\vec{s}} \cdot \hat{\vec{s}}'$.

The internal robot dynamics is independent of the spatial dimension. 
Therefore, the general structure of the Master equations, which describe the dynamics of MR, remains unchanged in three dimensions, but transport terms are adapted accordingly.
For MR controlled by NCS motif~$2$, these Master equations are given by
\begin{widetext}
\begin{subequations}
\label{eqn:master:3d}
  \begin{align}
	  \partial_t & \mathcal{P}_1 \! \left( \vec{r},\hat{\mathbf{s}},t \right)   =  - v_0 \: \! \hat{\mathbf{s}} \cdot  \nabla \mathcal{P}_1 + D_r \mathcal{L}[\mathcal{P}_i]
	  + D_{0} \Delta \mathcal{P}_1- \alpha[c] \mathcal{P}_1 + \gamma[c] \mathcal{P}_2 + \beta[c] \! \int \! d^3 \!\!\: s' \, g \! \left(  \hat{\mathbf{s}} | \hat{\mathbf{s}}' \right) \mathcal{P}_2 \! \left( \vec{r},\hat{\mathbf{s}}' \! ,t \right) \! , \\
	  \partial_t & \mathcal{P}_2\! \left( \vec{r},\hat{\mathbf{s}},t \right)   =  - v_0 \: \! \hat{\mathbf{s}} \cdot  \nabla \mathcal{P}_2 + D_r \mathcal{L}[\mathcal{P}_i] + D_{0} \Delta \mathcal{P}_2 - \Big ( \beta[c] + \gamma[c] \Big ) \mathcal{P}_2 + \alpha[c] \mathcal{P}_1 , 
  \end{align}
\end{subequations}
\end{widetext}
which is the analogue of~Eq.~\eqref{eqn:master:2d} for the corresponding two-dimensional case:~the convective term is replaced by its three dimensional equivalent, the reorientation distribution~$g\!\left( \varphi \right)$ is replaced by~$g \! \left(  \hat{\mathbf{s}} | \hat{\mathbf{s}}' \right)$ and the implementation of the director dynamics due to rotational noise has changed. 
The latter is determined in Cartesian coordinates by the operator 
\begin{align}
	\mathcal{L}[\mathcal{P}_i] = \partial_{s_\mu} \! \Big [ 2 s_{\mu} \mathcal{P}_i \Big ] \! + \partial_{s_\mu} \! \partial_{s_\nu} \! \Big [ \big( \delta_{\mu\nu} - s_\mu s_\nu \big ) \mathcal{P}_i  \Big ],
\end{align}
where a sum over~$\mu$ and~$\nu$ is implicit. 
This parametrization is entirely equivalent to the angular representation~[Eqs.~\eqref{eq:ang_dyn3d}], which can be verified by insertion of the parametrization of~$\hat{\vec{s}}$ via spherical coordinates~\cite{grossmann_anistropic_2015}.

In the diffusive limit, i.e.~if the external signal~$c \!\left( \vec{r} \right)$ varies weakly on spatial scales, which a robot traverses in between two reorientation events, a drift-diffusion approximation in the same spirit as in one and two spatial dimensions is feasible. 
The basic prerequisites of this derivation and its logic are analogous to the arguments put forward before;~technical subtleties are summarized in \red{Appendix}~\ref{app:E}. 
It turns out that solely the speed and the angular noise intensity are rescaled by numerical factors, which depend on the spatial dimension. 
The drift reads~[cf.~Eq.~\eqref{eqn:drift:2d}]
\begin{align}
	\label{eqn:drift:3d}
	\vec{f} \! \left( \vec{r} \right) &= - \Lambda_3 \nabla \Big( \! \ln \beta [c] \Big)
\end{align}
in three dimensions, where the prefactor~$\Lambda_3$ is structurally very similar to~$\Lambda_2$ determined by Eq.~\eqref{eqn:lambda_coeff} in two dimensions. 
Here, the prefactor~$\Lambda_3$ is determined by 
\begin{align}
	\! \Lambda_3 \!=\! \frac{v_0^2}{3} \! \cdot \! \frac{\beta_{+}\beta_{-} \! \left( 1 - \mathcal{G} \right)}{ \left ( \beta_{+} \! + \! \beta_{-} \right ) \! \cdot \! \Big [ \! \left( \beta_{+} \! + \! 2 D_r \right) \left( \beta_{-} \! + \! 2 D_r \right) - \beta_{+} \beta_{-} \mathcal{G} \, \Big ]} \, . \!
\end{align}
Along similar lines, only a few numerical factors are replaced in the expression for the effective diffusion coefficient, which reads in three dimensions as follows: 
\begin{align}
 	\label{eqn:diff:3d}
	D &= \, D_0 \\
	 & \!\!\! + \frac{v_0^2}{3} \! \cdot \! \frac{\beta_{+}^2 \!\!\;+\!\!\; \beta_{-}^2 \!\!\;+\!\!\; \beta_{+}\beta_{-}\!\left( 1 \!\!\;+\!\!\; \mathcal{G} \right) \!\!\;+\!\!\; 2 D_r \! \left( \beta_{+} \!\!\;+ \!\!\;\beta_{-} \right)}{ \left ( \beta_{+} \! + \! \beta_{-} \right ) \! \cdot \! \Big [ \! \left( \beta_{+} \! + \! 2 D_r \right) \left( \beta_{-} \! + \! 2 D_r \right) - \beta_{+} \beta_{-} \mathcal{G} \, \Big ] } . \nonumber
\end{align}
Just as in two dimensions, the parameter~$\mathcal{G}$ denotes the mean cosine of the angle between the directors before and after the reorientation event. 
In three dimensions, it may be expressed by 
\begin{align}
	\mathcal{G} = \mean{\cos \psi} \!=\!\! \int \! d^3 \!\!\: s \, \hat{\vec{s}} \cdot \hat{\vec{s}}' g \! \left( \hat{\vec{s}} | \hat{\vec{s}}' \right) \! = \! \int_{-1}^{1} \!\!d \! \left( \cos \psi \right) \!\!\: \mathscr{H} \! \left( \cos \psi \right) \! . 
\end{align}
The stationary probability density is thus determined by
\begin{align}
	\label{eqn:statpdf3d}
	P_s \! \left( \vec{r} \right) = \frac{\!\mathcal{N}}{\big (\beta \! \left[ c \right] \hspace{-0.02cm}\big )^{ \hspace{-0.05cm} \Lambda_3 / D}} . 
\end{align}
for adaptive MR controlled by NCS~motif~$2$.

\begin{figure}[tb]
 	\includegraphics[width=\columnwidth]{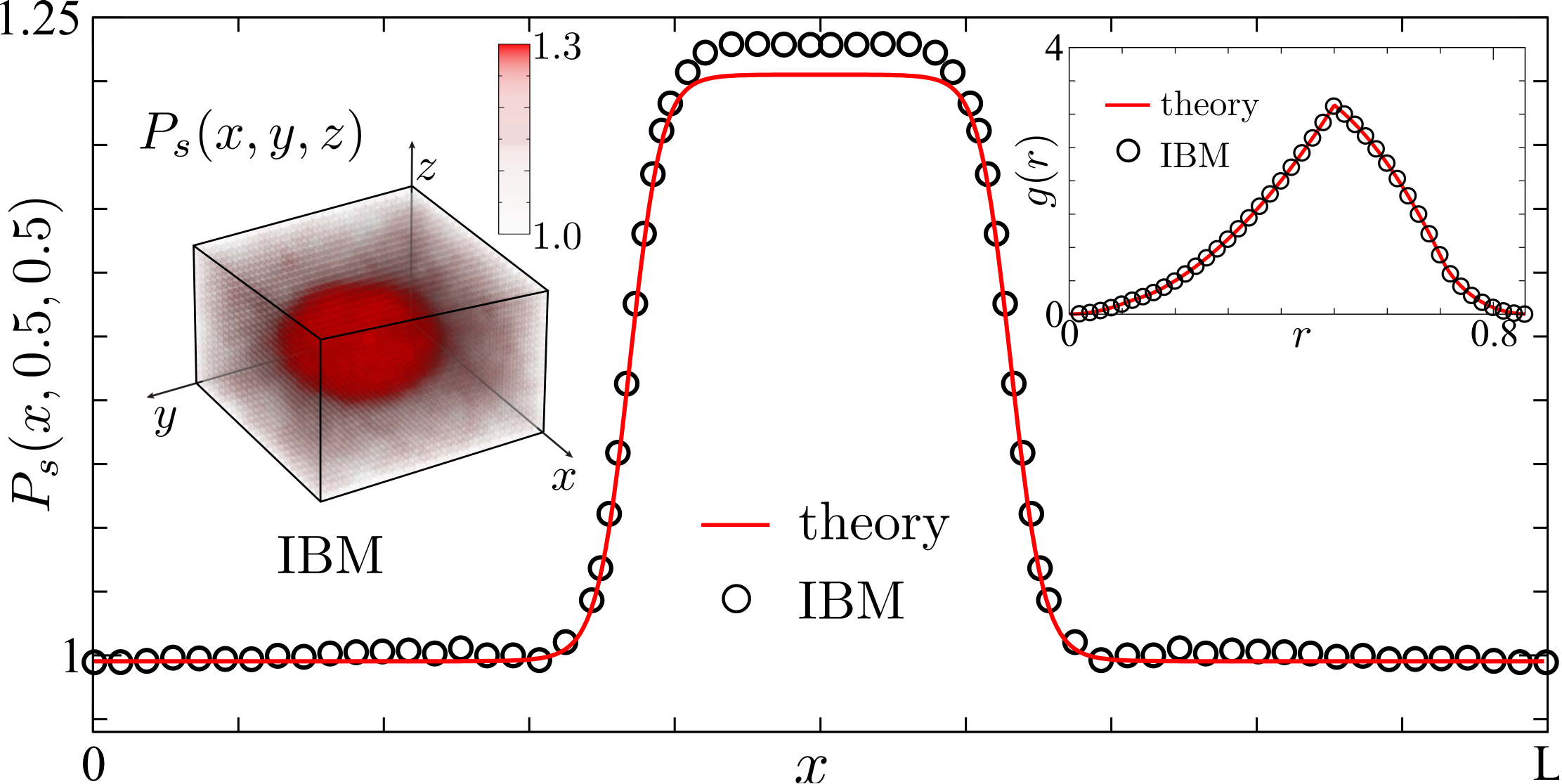}
	\caption{Comparison of the stationary probability density~$P_s \!\left( \vec{r} \right)$ of MR controlled by NCS motif~$2$ as obtained from individual-based model~(IBM) simulations and the corresponding drift-diffusion approximation, cf.~Eq.~\eqref{eqn:statpdf3d}, in three spatial dimensions. A Gaussian modulation is used as an external signal:~$c \!\left ( \vec{r} \right )= c_0 \! \left [ 1 \!\!\: + \!\!\: \varepsilon \exp \!\!\: \left( - \frac{\abs{\vec{r}-\vec{r}_0}^2}{2\sigma^2} \right) \! \right ]$. The data points in the main panel~(IBM) were reconstructed from the radial distribution function~$g(R) = \int \! d^3 r \, P_s \!\left( \vec{r} \right) \delta \! \left( R - \abs{\vec{r}-\vec{r}_{0}} \right)$ shown in the inset via division by the angular measure factor. The inset on the left represents a three-dimensional histogram of the position of MR, where the color code indicates the value of $P_s \!\left( \vec{r} \right)$. The main panel is a cut of~$P_s \!\left( \vec{r} \right)$ along~$\vec{r}=\left( x,1/2,1/2 \right)^T$. Parameters:~$c_0 = 1/2$, $\varepsilon = 2$, $\sigma = \sqrt{3/2}/10 \approx 0.12$, $\vec{r}_0 = \left( 1/2,1/2,1/2 \right)^T$. Further parameters as in Fig.~\ref{fig:2d}; reflecting boundary conditions have been used. }
  	\label{fig:3d}
\end{figure}

Notably, the simple rescaling of speed and rotational diffusion described above is not a particularity of the example under consideration, but it is generally the only quantitative difference of the transport properties of MR in two and three dimensions.
The proof of this result is sketched in \red{Appendix}~\ref{app:E}. In short, the behaviour of MR is qualitatively independent of the spatial dimension.

A comparison of individual-based model simulations and theoretical predictions in terms of the stationary probability density~$P_s \! \left( \vec{r} \right)$, as shown in Fig.~\ref{fig:3d}, serves as sanity check  that the analytically obtained transport coefficients, drift~[Eq.~\eqref{eqn:drift:3d}] and diffusion~[Eq.~\eqref{eqn:diff:3d}], provide a reasonable description  of the large-scale transport of MR  in the diffusive limit.

\section{Tests with a real robot}
\label{sec:robot}

\begin{figure}[tb]
  	\includegraphics[width=\columnwidth]{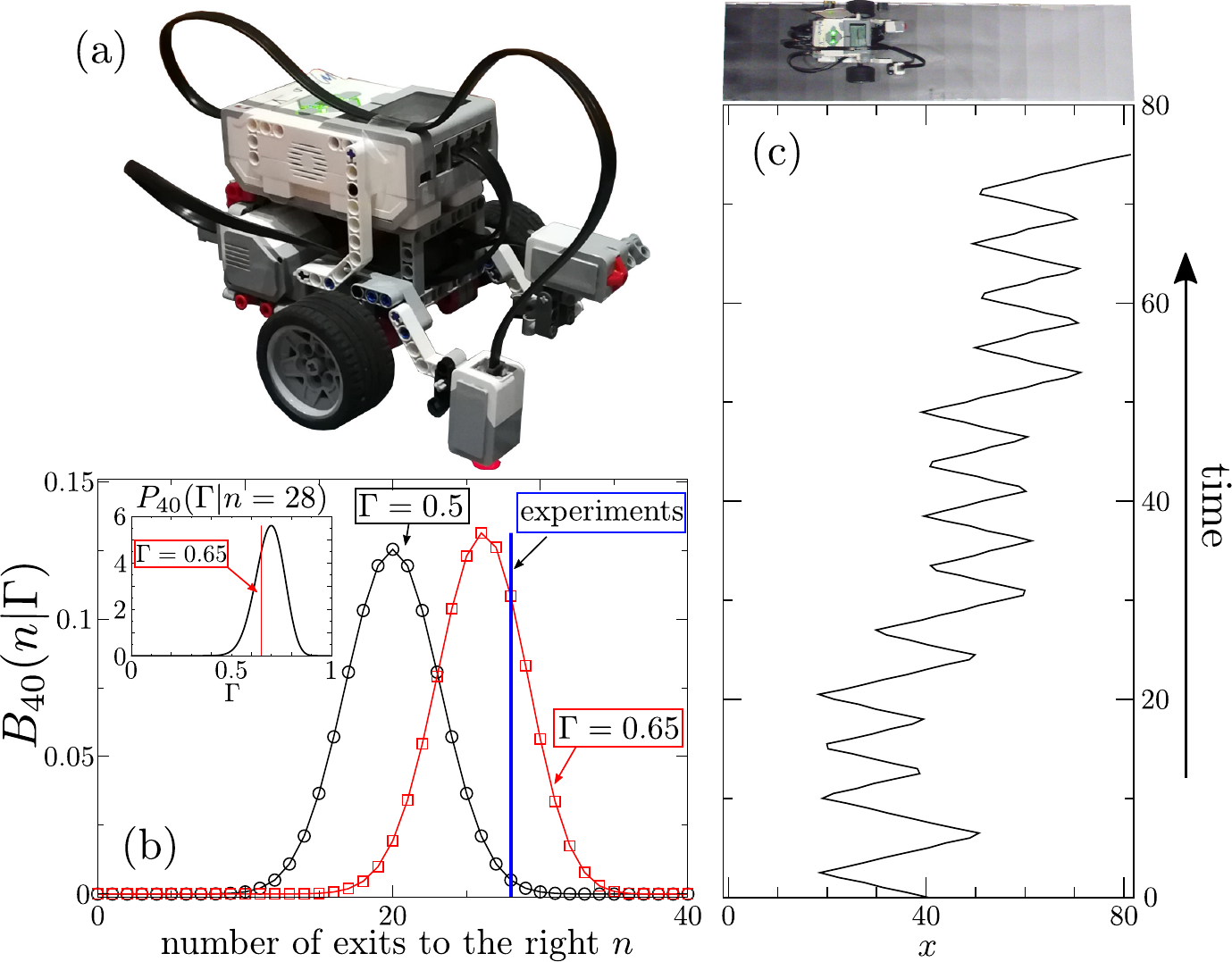}
	\caption{(a) A photo of the robot, a Lego Mindstorms~EV3. (b) Binomial distribution~$B_{40}(n|\Gamma)$ determining the probability that the robot leaves the system~$n$ times via the right boundary in~$40$ realizations of the experiment, whereby the robot was initially placed at the center of the system, given that the probability for the same event in one realization is~$\Gamma$, cf.~Eq.~\eqref{eqn:binomial}. The black circles correspond to an unbiased random walker~($\Gamma = 0.5$), and the red squares show the Binomial distribution for the theoretically predicted value~($\Gamma = 0.65$). The experimentally observed situation~--~in~$n=28$ cases the robot touched the right boundary first~--~is indicated by a vertical blue line. A representative trajectory is shown in panel~(c), on top of which the robot is depicted moving on the printed gray scale. The probability distribution~$P_{40}(\Gamma|n=28)$ for~$\Gamma$ given the experimental result~($n=28$), inferred from the Bayesian theorem~[Eq.~\eqref{eqn:bayesian}], is shown as an inset of panel~(b); the experimental observation is in line with the theoretical prediction. }
   	\label{fig:robot}
 \end{figure}

We tested the concepts developed before in practice by assembling a macroscopic robot that operates with NCS motif~$2$ as defined above. 
The robot~--~a Lego Mindstorms~EV3 shown in Fig.~\ref{fig:robot}\red{(a)}~--~was equipped with a single light sensor capable of reading light intensities, providing a signal~$\mathscr{S}$ in arbitrary units between~$0$ and~$100$ at the current position. 
A gray scale from black to white printed on paper~(total length:~$81\,\mbox{cm}$) was utilized as an external field~[cf.~the top panel of Fig.~\ref{fig:robot}\red{(c)}]. 
The robot possessed two synchronously steered motors in the front, each of which are connected to one wheel.
A metallic roller in the back of the robot served as stabilization. 
For simplicity, we focused on the one-dimensional scenario:~the robot was attached to a metallic rail to prevent turns, thereby ensuring straight trajectories.

Being a real-word system, the robot was naturally subjected to a series of fluctuations. 
Vibrations of the arm that connected the light sensor to the robot and, moreover, imperfections in the printed gray scale itself resulted in noisy measurements of the signal intensity~$\mathscr{S}$. 
Furthermore, imperfect rotations of the wheels imply varying step lengths and, hence, led effectively to noise in the particle position.

The robot was programed in the LabVIEW based Lego Mindstorms EV3-software. 
The basic flowchart of the algorithm is shown in Fig.~\ref{fig:ALGOS}.
The temporal update was composed of a streaming and a signal processing step that were repeated continuously.
The length of one streaming step was fixed to be~$2/3$ of the wheel perimeter, resulting in a step length of approximately~$11.7\, \mbox{cm}$. 
Afterwards, the signal strength~$\mathscr{S}$ was read from the sensor. 
Based on this measurement, the internal state was updated and, possibly, a reversal of the direction of rotation of the wheels could be triggered. 
The transition rates~$\alpha[c]$,~$\beta[c]$, and~$\gamma[c]$ were translated into probabilities ~$P_{\alpha}(\mathscr{S}) = b_{+}b_{-}/P_{\beta}(\mathscr{S})$,~$P_{\beta}(\mathscr{S}) = b_{+} \left ( b_{-}/b_{+} \right)^{ \mathscr{S} / 100 }$ and~$P_{\gamma}(\mathscr{S}) = b_{+} + b_{-} - P_{\alpha} - P_{\beta}$ for the corresponding transitions, where $b_{+}=1$ and $b_{-}=0.8$ were used.  
% 
% This sequence of steps is perpetuated subsequently. 

In the following, we aimed at testing the theoretical predictions at the level of exit probabilities.  
For this purpose, a single experimental run proceeded as follows:~it was monitored whether a robot which was initially placed in the middle of the experimental setup reached the left~(black) or right~(white) boundary of the system first. 
Once the robot touched one of the boundaries, the experiment was stopped and repeated. 
In total, $N=40$ realizations were recorded. 
In~$n=28$ cases, the robot left the system via the right boundary. 
A typical trajectory for an exit on the right boundary is displayed in Fig.~\ref{fig:robot}\red{(c)}.

Based on this experimental result, we first test the null hypothesis that the robot performed just an unbiased random walk. 
The exit probability on the right side of the system for a single experimental run should therefore be~$\Gamma = 0.5$. 
The probability to observe~$n$ exits on the right, given~$N$ realizations in total, is determined by the Binomial distribution
\begin{align}
	\label{eqn:binomial}
	B_N(n|\Gamma) = \binom{N}{n} \, \Gamma^n \left( 1-\Gamma \right)^{N-n}. 
\end{align}
This distribution is shown for~$N=40$ and~$\Gamma = 0.5$ in Fig.~\ref{fig:robot}\red{(b)} by black circles. 
The total probability to observe~$n=28$ exits to the right of the system or a more extreme result than this is determined by the tails of the Binomial distribution. 
It thereby constitutes the~$p$-value under the null hypothesis \textit{the robot performs an unbiased random walk}~\cite{freedman_statistics_1997}.
% %
% \begin{align}
% 	\sum_{n=0}^{12} B_N \! \left( n,\Gamma = 0.5  \right) + \sum_{n=28}^{40} B_N \! \left( n,\Gamma = 0.5  \right) \approx 0.017
% \end{align}
% %
In the case under consideration, we obtain a $p$-value of approximately~$0.017$. 
Accordingly, the null hypothesis may be discarded based on the standard significance level~$0.05$. 
In short, there is considerable statistical significance that the motion of the robot is biased due to the NCS at work.

The NCS was implemented such that the robot tended to move towards brighter areas in terms of the gray value and, thus, we expect the number of exits to the right to be larger than to the left. 
Simulations of the corresponding process predict that the probability to touch the right boundary first is~$\Gamma = 0.65(1)$. 
The Binomial distribution~$B_{40}\left( n|\Gamma \right)$ for this~$\Gamma$-value is represented by red squares in Fig.~\ref{fig:robot}\red{(b)}. 
The experimentally observed result~($n=28$) is indicated by a blue, vertical line. 
Apparently, the likelihood for the observed result given~$\Gamma = 0.65$ is higher than for the random walk:
\begin{align}
	B_{40} \!\left( n=28|\Gamma=0.65 \right) > B_{40} \!\left( n=28| \Gamma = 0.5 \right). 
\end{align}
Based on the Akaike information criterion~\cite{akaike_information_1998}, we infer that the theoretically predicted value~$\Gamma = 0.65$ is considerably more likely than the random walk hypothesis corresponding to~$\Gamma = 0.5$.

Finally, we specify the last statements regarding the likelihood of certain~$\Gamma$-values given the experimental observation.    
Using the Bayesian theorem~\cite{meyer_introductory_1965}, the following expression for the probability distribution~$P_{N}(\Gamma|n)$ for~$\Gamma$ given a certain number~$n$ of exits to the right out of~$N$ total experimental realizations is deduced:
\begin{align}
	\label{eqn:bayesian}
	P_{N}(\Gamma|n)	= \frac{B_{N} \!\left( n|\Gamma \right) \mathscr{P} \! \left( \Gamma \right)}{\int_{0}^{1} d \Gamma' \, B_{N} \!\left( n|\Gamma' \right) \mathscr{P} \!\left( \Gamma' \right) }. 
\end{align}
In the equation above, $\mathscr{P} \! \left( \Gamma \right)$ determines the prior knowledge~(before the experiment) about the probability distribution of~$\Gamma$. 
Here, we assume a uniform prior, i.e.~$\mathscr{P} \! \left( \Gamma \right) = 1$ for~$\Gamma \!\in\! [0,1]$ and zero otherwise. 
The probability distribution~$P_{40} \!\left( \Gamma| n=28 \right)$, which is relevant for the experimental result, is shown as an inset in Fig.~\ref{fig:robot}. 
Apparently, the distribution is shifted towards the right implying that there is a drift towards the right boundary as expected. 
The resulting distribution~$P_{40}(\Gamma|n=28)$ possesses a mean and standard deviation of~$\bar{\Gamma} = 0.69(7)$, which is well in line with the theoretical expectation.

In summary, the experimental results provide an empirical demonstration that the proposed navigation algorithm, which is simple to implement in a real robot, yields a directed, nontrivial motility response as predicted by theoretical considerations, which is, notably, robust with respect to fluctuations.

\section{Summary \& perspectives}

This study provides a solid proof-of-concept, including analytical derivations and a practical implementation, that it is possible to design robots that are capable of navigating through complex dynamical external fields in any spatial dimension~--~performing local measurements only~--~without making use of internal continuous variables to store previous measurements of the external field.  
The novel navigation strategies proposed and analyzed  
here are fundamentally different from bacterial chemotaxis (see Sec.~\ref{sec:1} for a detailed comparison).
It requires the robots to possess a minimum of two internal states  to exhibit non-trivial, persistent motility responses 
such as migration towards minima or maxima of the external field or even surfing at a desired field value in a complex, dynamical landscape. 
Transitions between these two internal states are dictated by a closed Markov chain, with transition rates 
that depend on the local, instantaneous value of the external field.
This implies that the internal dynamics of the robots is such that fixed points are excluded. 
In summary, we have shown here that robots with such a minimal navigation control system, 
where the internal state  can be stored in single Boolean variable, i.e. in 1 bit, are able to explore complex information landscapes.

Furthermore, we have shown that the proposed minimalistic navigation strategies can be efficiently implemented in real macroscopic 
robots. 
However, the main interest of conceiving navigation algorithms with limited memory storage capacity as the ones proposed here
is to pave the way to engineer miniature, micrometer-size robots in a near future. 
Miniaturizing robotic systems as the one used in Sec.~\ref{sec:robot} is a major technological challenge~\cite{churaman2012first,kim_microbiorobotics_2017}. 
Our study does not provide a recipe how to combine the existing micrometer-size actuators, sensors, and switches to produce the proposed Markovian robots, which is certainly beyond the scope of this basically theoretical work. 
Nevertheless, the developed concepts may serve as guiding principles to design autonomous, tiny robots capable of displaying complex motility behaviors by identifying the minimum requirements to navigate with limited memory storage capacity.

Finally, extensions of this initial study including more complex, biologically motivated motifs with a larger number of internal states may help to elucidate the navigation strategies of some microorganisms~\cite{nutsch2003signal,Harwood1990,Theves2013,hintsche_polar_2017}. 
Studying ensembles of interacting robots as those studied in~\cite{volpe2016}, operating with our navigation algorithms, is another promising research direction that may unveil cheap and efficient ways to obtain complex, self-organized collective behavior of autonomous, self-propelled agents.

%%%%%%%%%%%%%%%%%%%%%%%%%%%%%%%%%%%%%%%%%%%%%%%%%%%%%%%%%%%%%%%%%%%%%%%%%%%%%%%%%%%%%%
% appendix

\appendix

\section{Individual-based model~(IBM) simulations}
\label{app:A}

IBM simulations using~$N = 10^4$ robots were performed in order to check the validity of analytical approximations in the context of the reduction of the Master equation onto the Fokker-Planck dynamics. 
The state of a robot is characterized by three variables:~its position, its direction of active motion and the internal state depending on the motif under consideration. 
For motif~$1$, for example, the internal state can take the values~$1$ and~$2$. 
The spatial dynamics~[Eq.~(\ref{eqn:spat_dyn})] was solved by a stochastic Euler-Maruyama  method~\cite{gardiner_stochastic_2010} with a time step $\Delta t = 0.01$. 
The occurrence of reversal events are dictated by the internal variable as follows.
We implemented the evolution of the internal state using random numbers, which are uniformly distributed between~$0$ and~$1$. 
In each time step and for each robot, a transition in the motif was triggered if the random number is smaller than the product of the numerical time step~$\Delta t$ and the respective transition rate. 
Only one particular transition is accompanied by reversal, depicted by a dashed red arrow in each motif in Fig.~\ref{fig:scheme}. 
To get the stationary density distribution~$P_s(x)$, a histogram of robot positions was averaged over time. 
The total observation time was fixed to~$t_{obs} = 20\,000$ to ensure relaxation towards the stationary state.

\section{Numerical solution of Master equations} 
\label{app:B}

Individual-based simulations were validated by the direct integration of the Master equations~[Eqs.~(\ref{eqn:master_components_network_1_1}) and~(\ref{eqn:master_components_network_2_1})] corresponding to the NCS motif under consideration. 
Furthermore, the response of MR to a dynamic field gradient was performed numerically on the basis of the respective system of Master equations, cf.~Fig.~\ref{fig:tasks}. 
To solve those Master equations, a central finite difference discretization was employed in space and the temporal integration was performed using an explicit forward Euler algorithm~\cite{press1996numerical}. 
In particular, the spatial discretization~$\Delta x = 10^{-2}$ and temporal time step~$\Delta t = 10^{-3}$ was used in the context of Fig.~\ref{fig:tasks}.

\section{Drift-diffusion approximation in 1D} 
\label{app:C}

In the main text, the derivation of position-dependent drift and diffusion from the full set of Master equations is briefly sketched. 
In this paragraph, technical details of this derivation are presented in more detail for the one-dimensional case. 
Along with the general discussion of the principal ideas behind this derivation, NCS motif~$1$ is considered as an example. 
Effective Langevin equations for more complicated cases follow from the same procedure in a similar way.

Starting from the full Master equation for the probabilities~$P_{i}^{\pm} \! \left( x,t \right)$~[cf.~Eqs.~\eqref{eqn:master_components_network_2_1} for example], the change of variables~$P_i\! \left( x,t \right) = P_i^+\! \left( x,t \right) + P_i^-\! \left( x,t \right)$ and~$m_i\! \left( x,t \right) = P_i^+\! \left( x,t \right) - P_i^-\! \left( x,t \right)$ is performed as a first step, allowing us to recast the Master equation into two subgroups for the densities~$P_i \! \left( x,t \right)$ and the differences~$m_i \! \left( x,t \right)$~[cf.~Eqs.~\eqref{eqn:master_components_network_2_2}]: 
\begin{subequations}
 	\label{eqn:shortM}
	\begin{align}
		\partial_t P_i &= - v_0 \partial_x m_i + D_{0} \partial_{x}^2 P_i - \mathcal{Q}_{ij}[c]  P_j \, , \label{eqn:shortMa} \\
		\partial_t m_i &= - v_0 \partial_x P_i + D_{0} \partial_{x}^2 m_i - \mathcal{M}_{ij}[c]  m_j . \label{eqn:shortMb}
	\end{align}
\end{subequations}
Henceforward, Einsteins sum convention is used for the sake of compact notation. In the case of NCS~motif~$1$, the local transitions between the internal states are accounted for by the following matrices: 
\begin{align}
	\label{eqn:s:mat}
 	\!\mathcal{Q}[c] = \begin{pmatrix}
		\phantom{-}\alpha[c] &  -\beta[c] \\ 
		- \alpha[c] & \phantom{-}\beta[c]
	\end{pmatrix} \! , \;\;
	\mathcal{M}[c] = \begin{pmatrix}
		\phantom{-}\alpha[c] &  \phantom{-}\beta[c] \\ 
		- \alpha[c] & \phantom{-}\beta[c]
	\end{pmatrix} \! . \!\!
\end{align}

We begin the analysis with the dynamics of the differences~$m_i\! \left( x,t \right)$, given by Eq.~\eqref{eqn:shortMb}. 
The terms appearing are essentially of different types:~there is a~$m_i\! \left( x,t \right)$-independent source term proportional to the derivative of the densities~$P_i\! \left( x,t \right)$, diffusion of~$m_i\! \left( x,t \right)$ as well as local transitions. 
Now, diffusion is a slow process as compared to the exponential relaxation, which is described by the local transitions. 
Particularly, the real part of the eigenvalues~$\lambda_{\mathcal{M}}^{(\pm)}$ of the matrix~$\mathcal{M}$, given by
\begin{align}
	\lambda_{\mathcal{M}}^{(\pm)} = \frac{1}{2} \! \left[ \alpha + \beta \pm \sqrt{ \alpha^2 - 6 \alpha\beta + \beta^2} \, \right] \! ,
\end{align}
are strictly larger than zero for all positive rates. 
Accordingly, Eq.~(\ref{eqn:shortMb}) describes relaxation towards a stationary state. 
Assuming that this relaxation is a fast process, one may eliminate the variables~$m_i\! \left( x,t \right)$ adiabatically via~$\partial_t m_i \approx 0$. 
This yields the constitutive equation~$\!\!\!\!$
\begin{align}
	\mathcal{M}_{ij} m_j  \approx -v_0 \partial_x P_i + D_0 \partial_x^2 m_i . 
\end{align}
Since non of the eigenvalues of~$\mathcal{M}$ equals zero, the matrix~$\mathcal{M}$ is invertible:
\begin{align}
	\label{eqn:constMe}
	m_i \approx -v_0 \! \left \{\mathcal{M}^{-1} \right \}_{ij} \partial_x P_j + D_0 \! \left \{\mathcal{M}^{-1} \right \}_{ij} \partial_x^2 m_j . 
\end{align}
A closed expression for~$m_i\! \left( x,t \right)$ in terms of~$P_i\! \left( x,t \right)$ can be found by recursive reinsertion on the right hand side. 
With regard to the objective of this derivation, we turn now, however, to the~$P_i\! \left( x,t \right)$-dynamics~[Eq.~\eqref{eqn:shortMa}]. 
Notably, we want to obtain a closed equation up to second order in spatial derivatives. 
The~$P_i \! \left( x,t \right)$-dynamics is driven by first order derivatives of~$m_i\! \left( x,t \right)$, which are, in turn, proportional to derivatives of~$P_i\! \left( x,t \right)$ to lowest order. 
Hence, it is sufficient to truncate the recursion~[Eq.~(\ref{eqn:constMe})] at the lowest order in spatial derivatives: 
\begin{align}
	\label{eqnS:constM1d}
	m_i \approx - v_0 \! \left \{\mathcal{M}^{-1} \right \}_{ij} \partial_x P_j. 
\end{align}
Accordingly, we obtain the following expression for the dynamics of~$P_i \! \left( x,t \right)$ as an intermediate result: 
\begin{align}
	\label{eqn:redP}
	\partial_t P_i = \partial_x \bigg[ \Big (v_{0}^2 \! \left \{\mathcal{M}^{-1} \right \}_{ij} \! + D_0 \delta_{ij} \Big ) \partial_x P_j \bigg] - \mathcal{Q}_{ij}[c] P_j . 
\end{align}
The dynamics is a combination of position-dependent diffusion as well as local transitions between the different internal states.

In contrast to the matrix~$\mathcal{M}$, the matrix~$\mathcal{Q}$ possesses always one eigenvalue which equals zero.
It results from the fact that the robot must be in one of its internal state. 
This \textit{conservation law} implies a zero-eigenmode corresponding the slow dynamics of the total, conserved density~$P \! \left( x,t \right) \! =  \! \sum_{i} P_i \! \left( x,t \right)$. 
Other eigenvalues are positive implying the existence of additional fast modes [notice the minus sign in front of $\mathcal{Q}_{ij}$ in Eq.~(\ref{eqn:redP})]. 
The two eigenvalues of the matrix~$\mathcal{Q}$ read $\lambda_{\mathcal{Q}}^{(1)} \!=\! \alpha\!+\!\beta$ and $\lambda_{\mathcal{Q}}^{(2)} \!=\! 0$ for NCS motif~$1$ for example, cf.~Eq.~\eqref{eqn:s:mat}. 
To lowest order in spatial gradients, the adiabatic elimination of the fast mode reveals that~$P_i \! \left( x,t \right)$ must be an element of the kernel of~$\mathcal{Q}$:
\begin{align}
	\label{eqn:lE}
	\mathcal{Q}_{ij} P_j \approx 0. 
\end{align}
Physically, this reflects the assumption of local equilibrium implying that the local transitions are much faster compared to the motion of robots such that the local distribution of robots among the different internal states is equalized. 
This is in line with the general scope of this work:~the external signal is weakly space-dependent, i.e.~the field~$c(x)$ varies on scales which are much larger than the mean distance~$l_b \!=\! v_0 \tau$, which a robot travels in between two reorientation events that occur at an average rate~$\tau^{-1}$; in short, only local measurements of the external signal are feasible. 
Consequently, the~$P_i \! \left( x,t \right)$-dynamics can relax locally faster than the overall density distribution on scales larger than~$l_b$.

There is a nontrivial solution~$P_i \! \left( x,t \right) \!=\! P \! \left( x,t \right) V_i[c]$ to Eq.~(\ref{eqn:lE}) since the matrix~$\mathcal{Q}$ is not invertible~\footnote{The existence of a unique, nontrivial, stationary solution for this type of Master equation, which reflects the transition dynamics in between the internal states, is ensured in general~\cite{kampen_stochastic_2011}. }. 
This solution is, however, unique due to the normalization condition~$P\! \left( x,t \right) = \sum_i P_i\! \left( x,t \right)$ which implies necessarily that the sum of the components of the vector~$\mathbf{V}[c]$ equals one.  
For the example of NCS motif~$1$ considered above, we obtain
\begin{align}
	\mathbf{V}[c] =  \frac{1}{\alpha[c]+\beta[c]} 
	\begin{pmatrix}
		\beta[c] \\ \alpha[c]
	\end{pmatrix} \! . 
\end{align}
Inserting this closure into the reduced~$P_i \! \left( x,t \right)$-equation~[Eq.~(\ref{eqn:redP})] and subsequent summation over all components 
yields eventually the following closed equation for the total density: 
\begin{align}
	\! \! \partial_t P = \sum_{i,j} \partial_x \! \left \{ \! \Big (v_{0}^2 \! \left \{ \mathcal{M}^{-1} \right \}_{ij} \! + D_0 \delta_{ij} \Big )  \partial_x \big [ P \! \left( x,t \right) \! V_j[c] \big ] \! \right \} \! . \!\!\!\!
\end{align}
In order to read of the mean drift~$f(x)$ as well as the position-dependent diffusion coefficient~$D(x)$, terms have to be rearranged to meet the structure of a Fokker-Planck equation in Ito form~\cite{gardiner_stochastic_2010}: 
\begin{align*}
	\label{eqn:structFP1}
	\partial_t P \! \left( x,t \right) = - \partial_x \Big[ f(x) P \! \left( x,t \right) \Big ] \!  + \partial_x^2 \Big [ D(x) P \! \left( x,t \right) \Big ] .
\end{align*}
From this Fokker-Planck equation, which defines~$f(x)$ and~$D(x)$ unambiguously, we read off 
	\begin{align*}
		f(x) &= v_0^2 \sum_{i,j} \Big ( \partial_x \! \left \{ \! \mathcal{M}^{-1} \right \}_{ij} \! \Big ) V_j[c] , \\ 
		D(x) &= D_0 + v_0^2 \sum_{i,j} \left \{ \mathcal{M}^{-1} \right \}_{ij} \! V_j[c] .
		\end{align*}
Inserting the inverse of~$\mathcal{M}$ for NCS motif~$1$, 
\begin{align}
	\mathcal{M}^{-1} = \frac{1}{2} \begin{pmatrix}
		1/\alpha[c] & -1/\alpha [c]\\ 
		1/\beta[c]  & \phantom{-} 1/\beta[c]
	\end{pmatrix} \! ,
\end{align}
yields eventually the expressions which are given in the main text~[c.f.~Eqs.~\eqref{eqn:master_components_network_2_4}].

\section{Drift-diffusion approximation in 2D} 
\label{app:D}

The extension of the drift-diffusion approximation to two (or higher) spatial dimensions is straightforward on the basis of the previously described derivation of effective Langevin equations in one dimension. 
The conceptual basis is unchanged:~at first, a closed expression for the probability densities~$P_i \! \left( \vec{r},t \right)$ to find a robot at a certain position~$\vec{r}$ at time~$t$ is derived by adiabatic elimination of fast order parameters and, in a second step, this set of equations is reduced to the total density assuming local equilibrium. 
There are two technical complications which need particular attention. 
In dimensions larger than one, there are two vector spaces that need to be distinguished:~the physical space which robots move in as well as the space of internal states. 
As before, we use Latin indices to label internal states~($\mathcal{P}_i$) and, from now on, vectorial notation is used to indicate contractions with respect to the vector space of spatial coordinates~($\vec{r}$). 
Further, it turns out to be crucial to identify the correct generalizations of the central quantities of interest in one dimension, namely densities~$P_i(x,t)$ and differences~$m_i(x,t)$, for the two-dimensional situation.

Starting point of the derivation is the Master equation~[cf.~Eq.~\eqref{eqn:master:2d}] for the probability densities~$\mathcal{P}_i \! \left( \vec{r},\varphi,t \right)$ to find a robot in state~$i$ at position~$\vec{r}$ moving into the direction~$\varphi$ at time~$t$. 
In general, the dynamics is of the form
\begin{align}
	\label{eqn:master:2ds:SUPP}
	\partial_t & \mathcal{P}_i \! \left( \vec{r},\varphi,t \right)  \! =\!  - v_0 \hat{\mathbf{s}}[\varphi] \! \cdot \! \nabla \mathcal{P}_i + D_r \partial_\varphi^2 \mathcal{P}_i + D_{0} \Delta \mathcal{P}_i \!\!\\
	& - \! \bar{\gamma}_i[c] \mathcal{P}_i  + \! \sum_{j} \sum_{k=1}^{n_c} \gamma^{(k)}_{ij} [c] \!\! \int_{\!-\pi}^{\pi} \!\! d\varphi' \, g^{(k)}_{ij} \! \left( \varphi - \varphi' \right) \! \mathcal{P}_j \! \left( \vec{r},\varphi' \! ,t \right) \! . \nonumber 
\end{align}
The terms in the first line describe the motility of robots:~active motion along the director~$\hat{\mathbf{s}}[\varphi(t)] = (\cos \varphi, \sin \varphi)$, rotational diffusion due to spatial heterogeneities or fluctuations of the active force~\cite{mikhailov_self_1997,peruani_self_2007,romanczuk_brownian_2011,chepizhko2013} giving rise to a diffusion term with respect to the polar angle~$\varphi$, and isotropic diffusion. 
Stochastic transitions from one internal state to another are accounted for by the second line. 
The total rate at which the state~$i$ is left is determined by the rate
\begin{align}
	\bar{\gamma}_i[c] = \sum_{j} \sum_{k=1}^{n_c} \gamma^{(k)}_{ji}[c]. 
\end{align}
The transition rates ~$\gamma_{ij}^{(k)}$ denote the probability per unit time for a transition from~$j$ to state~$i$ via the~$k$-ths channel~(number of channels:~$n_c$). 
The~$\boldsymbol{\gamma}$-matrices are the immediate mathematical representation of the NCS motif under consideration. 
In the case of NCS motif~$1$, that was used as an example before, there is only one channel for each transition such that the $\boldsymbol{\gamma}$-matrix reads
\begin{align}
	\label{eqn:s:nCS2}
	\boldsymbol{\gamma}^{(1)} = \begin{pmatrix}
		0 & \beta[c] \\
		\alpha[c] & 0
	\end{pmatrix} \! .
\end{align}
For NCS motif~$2$, in contrast, two~$\boldsymbol{\gamma}$-matrices have to be introduced since there are two potential transitions from states~$2$ to~$1$, cf.~Fig.~\ref{fig:scheme}, one of which is accompanied by a reorientation whereas the other one is not:
\begin{align}
	\boldsymbol{\gamma}^{(1)} = \begin{pmatrix}
		0 & \gamma[c] \\
		\alpha[c] & 0
	\end{pmatrix} \! , \; \;
	\boldsymbol{\gamma}^{(2)} = \begin{pmatrix}
		0 & \beta [c] \\
		0 & 0
	\end{pmatrix} \! . 
\end{align}
Reorientations in space upon transitions are accounted for by the probability distributions~$g_{ij}^{(k)}(\varphi)$.

Now, drift and diffusion properties of MR in two dimensions are derived along the line of arguments which was introduced in the previous paragraph for the one-dimensional case.

By integration of Eq.~\eqref{eqn:master:2ds:SUPP} over all angles~$\varphi$, we obtain the dynamics of the probability densities~$P_i \! \left( \vec{r},t \right)$ to find a robot at position~$\vec{r}$ at time~$t$, independent of its direction of motion: 
\begin{align}
	\label{eqn:densdyn:2d}
	\partial_t P_i \! \left( \vec{r},t \right) = - v_0 \nabla \cdot \vec{m}_i + D_0 \Delta P_i - \mathcal{Q}_{ij} [c] P_j. 
\end{align}
This equation is structurally equivalent to Eq.~\eqref{eqn:shortMa} in one dimension. 
The elements of the~$\mathcal{Q}$-matrix read in general
\begin{align}
	\label{eqn:s:defQ}
	\mathcal{Q}_{ij} = - \sum_{k=1}^{n_c} \left [ \gamma_{ij}^{(k)} - \delta_{ij} \sum_{l} \gamma_{lj}^{(k)} \right ] \! .
\end{align}
One may check easily that this definition of~$\mathcal{Q}$ yields consistently the known expression for NCS motif~$1$~[Eq.~\eqref{eqn:s:mat}], for example, if the corresponding~$\boldsymbol{\gamma}$-matrix~[Eq.~\eqref{eqn:s:nCS2}] is inserted. 
Naturally, those local terms in Eq.~\eqref{eqn:densdyn:2d} corresponding to the internal robot dynamics remain unchanged since they are independent of the spatial dimension. 
Consistently, only the terms related to transport in space are altered in two dimensions as compared to the 1D scenario. 

Replacing the density differences $m_i$ between  left- and right-moving robots in state $i$, used to analyze the 1D scenario and determining density transport, in two dimensions we make use of the  \textit{local order parameter} 
\begin{align}
	\mathbf{m}_i \! \left( \vec{r},t \right) \! = \!\! \int \!\! d \varphi \,  \hat{\mathbf{s}} \! \left[ \varphi \right] \! \mathcal{P}_i \! \left( \vec{r},\varphi,t \right)
	\!=\! \!\int \!\!d \varphi 
	\begin{pmatrix}
		\cos \varphi \\ \sin \varphi
	\end{pmatrix} \! 
	\mathcal{P}_i \! \left( \vec{r},\varphi,t \right) \, ,
\end{align}
which appears in Eq.~\eqref{eqn:densdyn:2d} and that when multiplied by~$v_0$ provides the flux due to active self-propulsion. 
It is, further, the first Fourier mode of the probability distribution function~$\mathcal{P}_i \! \left( \vec{r},\varphi,t \right)$. 
In general, the dynamics of the fields~$\vec{m}_{i} \! \left( \vec{r},t \right)$ is coupled to higher order Fourier modes of the probability densities~$\mathcal{P}_{i} \! \left( \vec{r},\varphi,t \right)$ giving rise to an infinite hierarchy. 
However, we can make use of the fact that the dynamics of higher order Fourier modes is fast, i.e., their dynamics is slaved~\cite{haken_synergetics_2004} to the density in the long time limit thus allowing for their adiabatic elimination~\footnote{For a detailed account on mode reduction in several active motion models, see~\cite{bertin_boltzmann_2006,grossmann_self-propelled_2013,peruani2017}.}. 
Similar arguments as in the one-dimensional case apply:~since we aim at a reduction of the dynamics to a drift-diffusion equation, it is sufficient to calculate the dynamics of the mean local orientations~$\vec{m}_{i} \! \left( \vec{r},t \right)$ to first order in density gradients. 
Accordingly, we derive the following dynamics of~$\vec{m}_{i} \! \left( \vec{r},t \right)$ by multiplication of the full Master equation with~$\hat{\mathbf{s}} \! \left[ \varphi \right]$ and subsequent integration over the polar angle~$\varphi$:
\begin{align}
	\label{eqn:mdyn:2d}
	\partial_t \vec{m}_i \! \left( \vec{r},t \right) \simeq - \frac{v_0}{2} \nabla P_i + D_0 \Delta \vec{m}_i - \mathcal{M}_{ij}[c] \vec{m}_{j}. 
\end{align}
This is a straightforward generalization of Eq.~\eqref{eqn:shortMb}. In two dimensions, the matrix elements of the local dynamics read
\begin{align}
	\label{app:eqn:mij} 
 	\mathcal{M}_{ij} = D_r \delta_{ij} - \sum_{k=1}^{n_c} \left [ \gamma_{ij}^{(k)} \! \mean{\cos \varphi}_{ij}^{(k)} - \delta_{ij} \! \sum_{l} \gamma_{lj}^{(k)} \right ] \!. 
\end{align}
In two dimensions, they depend on the mean cosine of the reorientation distributions~\footnote{It was silently assumed that the reorientation distributions~$g_{ij}^{(k)}(\varphi)$ are symmetric:~$g_{ij}^{(k)}(-\varphi) = g_{ij}^{(k)}(\varphi)$. }
\begin{align}
	\mean{\cos \varphi}_{ij}^{(k)} = \int_{\! -\pi}^{\pi} \! d\varphi \, \cos \varphi \, g_{ij}^{(k)} \! \left( \varphi \right). 
\end{align}
In the case of reversal for NCS motif~$1$, $g_{12}^{(1)} \! \left( \varphi \right) = \delta \! \left( \varphi - \pi \right)$ is the only nontrivial element. 
It implies~$\mean{\cos \varphi}_{12}^{(1)} = -1$. 
In combination with the corresponding~$\boldsymbol{\gamma}$-matrix~[Eq.~\eqref{eqn:s:nCS2}], the matrix~$\mathcal{M}[c]$ reduces again exactly to the familiar one-dimensional result, cf.~Eq.~\eqref{eqn:s:mat}.

The reduction of Eqs.~\eqref{eqn:densdyn:2d} and~\eqref{eqn:mdyn:2d} onto the density follows the procedure which was outline in the previous paragraph for one spatial dimension. 
Adiabatic elimination~($\partial_t \vec{m}_i \approx 0$) of the fields~$\vec{m}_i$ yields
\begin{align}
	\vec{m}_i \approx - \frac{v_0}{2} \! \left \{ \mathcal{M}^{-1} \right \}_{ij} \! \nabla P_j
\end{align}
to lowest order in density gradients~[cf.~Eq.~\eqref{eqnS:constM1d}] and, thus, one obtains the following reduced density dynamics via insertion into Eq.~\eqref{eqn:densdyn:2d}: 
\begin{align}
	\label{eqn:tmp:P}
	\partial_t P_i \! \left( \vec{r},t \right) \simeq \nabla \! \cdot \! \left [ \left ( \!
		\frac{v_0^2}{2} \! \left \{ \mathcal{M}^{-1}\right \}_{ij}  \! + \! D_0 \delta_{ij} \right ) \! \nabla P_j
	\right ] \! - \mathcal{Q}_{ij}[c] P_j. 
\end{align}
This is the 2D-analogue to Eq.~\eqref{eqn:redP}. 
In the diffusive limit, the fields~$P_i$ have to lie in the kernel of the matrix~$\mathcal{Q}[c]$, i.e.~$\mathcal{Q}_{ij}[c] P_j = 0$. 
We normalize such that~$P_i \! \left( x,t \right) = P(x,t) V_i[c]$ implying~$\sum_i V_i[c] = 1$ and~$\mathcal{Q}_{ij}[c] V_j = 0$. 
Inserting this ansatz into Eq.~\eqref{eqn:tmp:P} yields the preliminary Fokker-Planck equation 
\begin{align}
	\label{eqn:prov:2d:FP} 
	\partial_t P \! \left( \vec{r},t \right) \simeq \sum_{i,j} \nabla \cdot \left \{ \! \left (
		\frac{v_0^2}{2} \! \left \{ \mathcal{M}^{-1} \right \}_{ij} + D_0 \delta_{ij} \right ) \! \nabla \big [ P V_j [c] \big ] \! \right \} \! .
\end{align}
We define the force~$\mathbf{f} \! \left( \vec{r} \right)$ and the position-dependent diffusion~$D \! \left( \vec{r} \right)$ analogous to the one-dimensional case~[Eq.~\eqref{eqn:structFP1}]: 
\begin{align}
	\partial_t P \! \left( \vec{r},t \right) = - \nabla \! \cdot \! \Big[ \vec{f} \! \left( \vec{r} \right) \! P \! \left( \vec{r},t \right) \! \Big] \! + \Delta \Big[ D \! \left( \vec{r} \right) \! P \! \left( \vec{r},t \right) \! \Big] .
\end{align}
The comparison to Eq.~\eqref{eqn:prov:2d:FP} eventually yields the final expressions for the drift and local diffusion coefficient: 
\begin{subequations}
\begin{align}
	\vec{f} \! \left( \vec{r} \right) &=  \frac{v_0^2}{2} \sum_{i,j} \left( \nabla \! \left \{ \mathcal{M}^{-1}\right \}_{ij}  \right) \! V_j[c] , \\
	D \! \left( \vec{r} \right) & = D_0 + \frac{v_0^2}{2} \sum_{i,j} \left \{\mathcal{M}^{-1}\right \}_{ij} \!  V_j[c] . 
\end{align}
\end{subequations}

\section{Drift-diffusion approximation in 3D} 
\label{app:E}

In this section, we briefly summarize some technical particularities of the drift-diffusion approximation in three dimensions. 
The general discussion follows closely the procedure outlined in \red{Appendix}~\ref{app:D}. 
Starting point is the general Master equation
\begin{align}
	\label{app:eqn:master3d}
	\partial_t & \mathcal{P}_i \! \left( \vec{r},\hat{\vec{s}},t \right)  \! =\!  - v_0 \!\: \hat{\mathbf{s}} \! \cdot \! \nabla \mathcal{P}_i + D_r \mathcal{L}[\mathcal{P}_i] + D_{0} \Delta \mathcal{P}_i \!\!\\
	& - \! \bar{\gamma}_i[c] \mathcal{P}_i  + \! \sum_{j} \sum_{k=1}^{n_c} \gamma^{(k)}_{ij} [c] \!\! \int \! d^3 \!\!\: s' \, g^{(k)}_{ij} \! \left( \hat{\vec{s}}|\hat{\vec{s}}' \right) \! \mathcal{P}_j \! \left( \vec{r},\hat{\vec{s}}' \! ,t \right) \! . \nonumber 
\end{align}
The transition rates~$\gamma^{(k)}_{ij}[c]$ denote, as before, the probability per unit time for a transition from state~$j$ to state~$i$ via the~$k$-ths channel. 
Such transitions may be accompanied by a transition from an orientation~$\hat{\vec{s}}'$ to~$\hat{\vec{s}}$ which is accounted for by the transition probability density~$g^{(k)}_{ij} \! \left( \hat{\vec{s}}|\hat{\vec{s}}' \right)$. 
The continuous, stochastic rotational dynamics of the director~$\hat{\vec{s}}$ in three dimensions is accounted for by the operator
\begin{align}
	\mathcal{L}[\mathcal{P}_i] = \partial_{s_\mu} \! \Big [ 2 s_{\mu} \mathcal{P}_i \Big ] \! + \partial_{s_\mu} \! \partial_{s_\nu} \! \Big [ \big( \delta_{\mu\nu} - s_\mu s_\nu \big ) \mathcal{P}_i  \Big ],
\end{align}
where a sum over~$\mu$ and~$\nu$ is implicit. 
This Cartesian representation of the director dynamics is simpler to handle in terms of analytical calculations as compared to a parametrization in terms of spherical coordinates~(cf.~Eq.~\eqref{eq:ang_dyn3d} and Refs.~\cite{grossmann_anistropic_2015,hintsche_polar_2017}).

We point out that there are two vector spaces which have to be distinguished in the following:~the physical space which robots move in~(three dimensional) and the space of internal states. 
To avoid confusion, the components of the former are denoted by Greek indices, whereas the latter are indicated by Latin indices.

The derivation of the drift-diffusion approximation starts from the temporal evolution of the densities
\begin{align}
	P_i \!\left( \vec{r},t \right) = \int \! d^3 \!\!\: s \, \mathcal{P}_i \!\left( \vec{r},\hat{\vec{s}},t \right) \!,
\end{align}
which is obtained from the Master equation~\eqref{app:eqn:master3d} by integration over all orientations of the director yielding
\begin{align}
	\partial_t P_i \! \left( \vec{r},t \right) = - v_0 \nabla \cdot \vec{m}_i + D_0 \Delta P_i - \mathcal{Q}_{ij} [c] P_j. 
\end{align}
Just as in two dimensions, the matrix elements of the matrix~$\mathcal{Q}$ are determined by
\begin{align}
	\mathcal{Q}_{ij} = - \sum_{k=1}^{n_c} \left [ \gamma_{ij}^{(k)} - \delta_{ij} \sum_{l} \gamma_{lj}^{(k)} \right ] \! .
\end{align}
We keep in mind that we will assume local equilibrium throughout, i.e.~the probability to find a robot in a certain internal state is determined by~$P_i \!\left( \vec{r},t \right) = V_i[c] P \!\left( \vec{r},t \right)$ such that~$\mathcal{Q}_{ij} V_j = 0$ and~$\sum_i V_i = 1$.

In three dimensions, the flux is determined by the local order parameter
\begin{align}
	\vec{m}_i \!\left( \vec{r},t \right) = \! \int \! d^3 \!\!\: s \, \hat{\vec{s}} \, \mathcal{P}_i \!\left( \vec{r},\hat{\vec{s}},t \right) \! .
\end{align}
The dynamics of~$\vec{m}_i$ is, in turn, obtained by multiplication of the Master equation~\eqref{app:eqn:master3d} by~$\hat{\vec{s}}$ and subsequent integration: 
\begin{align}
	\label{app:eqn:mi3d}
	\partial_t \vec{m}_i \! \left( \vec{r},t \right) = - v_0 \nabla \cdot \mathcal{T}_{i} + D_0 \Delta \vec{m}_i - \mathcal{M}_{ij}[c] \vec{m}_{j}. 
\end{align}
The matrix elements
\begin{align}
 	\! \mathcal{M}_{ij} = 2 D_r \delta_{ij} \!\!\:-\!\!\: \sum_{k=1}^{n_c} \left [ \gamma_{ij}^{(k)} \! \mean{\cos \psi}_{ij}^{(k)} \! - \delta_{ij} \! \sum_{l} \gamma_{lj}^{(k)} \right ]
\end{align}
differ from their two-dimensional counterpart~[cf.~Eq.~\eqref{app:eqn:mij}] by just a factor of two in front of the angular noise intensity. 
The relevant parameter which accounts for the reorientation is the mean cosine of the angle between the directors just right before and after a reorientation, defined by
\begin{align}
	\mean{\cos \psi}_{ij}^{(k)} = \! \int \! d^3 \!\!\: s \, \hat{\vec{s}} \cdot \hat{\vec{s}}' g_{ij}^{(k)} \!\left( \hat{\vec{s}} | \hat{\vec{s}}' \right) \! .
\end{align}
Due to the Cartesian parametrization of the director dynamics, a new term involving the symmetric tensor 
\begin{align}
	\big \{\mathcal{T}_i \big \}_{\mu \nu} =  \int \! d^3 \!\!\: s \, s_{\mu} s_{\nu} \:\! \mathcal{P}_i \!\left( \vec{r},\hat{\vec{s}},t \right)
\end{align}
appears in Eq.~\eqref{app:eqn:mi3d}. 
In order to derive an effective transport equation for the density~$P \!\left( \vec{r},t \right) = \sum_{i} P_i \! \left( \vec{r},t \right)$, a closure relation for the tensors~$\mathcal{T}_i$ has to be found.
Since we aim at reducing the density dynamics to a Fokker-Planck equation valid in the diffusive limit, it is possible to neglect  spatial derivatives in the dynamics of~$\mathcal{T}_i$ to lowest order: 
\begin{align}
	\partial_t \big \{\mathcal{T}_i \big \}_{\mu \nu} \approx \delta_{\mu \nu} \Omega_{ij} P_j -  \Xi_{ij} \big \{ \mathcal{T}_j \big \}_{\mu \nu} \:\! . 
\end{align}
This equation involves the matrices
\begin{align}
	\Omega_{ij} = 2 D_r \:\! \delta_{ij} + \sum_{k=1}^{n_c} \gamma_{ij}^{(k)}[c] \cdot \! \frac{1 - \mean{\cos^2 \psi}_{ij}^{(k)}}{2} 
\end{align}
and
\begin{align}
	\Xi_{ij} \!=\! \Big ( 6 D_r + \bar{\gamma}_i[c] \Big ) \:\! \delta_{ij} \!\!\:-\!\!\: \sum_{k=1}^{n_c} \!\!\: \left [ \gamma_{ij}^{(k)}[c] \cdot \! \frac{3 \!\!\:\mean{\cos^2 \psi}_{ij}^{(k)} \!\!\:-\!\!\: 1}{2} \right ] \! .
\end{align}
The tensor~$\mathcal{T}_i$ may be expressed as a function of the densities~$P_i \!\left( \vec{r},t \right)$ via adiabatic elimination,~$\partial_t \big \{\mathcal{T}_i \big \}_{\mu \nu} \approx 0$. 
In the state of local equilibrium, where~$P_i = V_i P$ and~$\mathcal{Q}_{ij}V_j = 0$, the rather complicated expressions above take a rather simple form, as can be verified via direct calculation: 
\begin{align}
	\big \{\mathcal{T}_i \big \}_{\mu \nu} = \frac{\delta_{\mu\nu}}{3} \, V_i[c] P. 
\end{align}
Reinsertion of this solution into the dynamics of~$\vec{m}_i$~[Eq.~\eqref{app:eqn:mi3d}] yields the familiar equation
\begin{align}
	\partial_t \vec{m}_i \! \left( \vec{r},t \right) \simeq - \frac{v_0}{3} \!\:\nabla P_i + D_0 \Delta \vec{m}_i - \mathcal{M}_{ij}[c] \vec{m}_{j}
\end{align}
which is structurally identical to~Eqs.~\eqref{eqn:shortMb} and~\eqref{eqn:mdyn:2d}.
Only the speed has been rescaled by the spatial dimensionality.

The remaining part of the calculation follows therefore exactly the same steps as in two spatial dimensions. 
Accordingly, the drift and diffusion for MR in three spatial dimensions are determined by 
\begin{subequations}
\begin{align}
	\vec{f} \! \left( \vec{r} \right) &=  \frac{v_0^2}{3} \sum_{i,j} \left( \nabla \! \left \{ \mathcal{M}^{-1}\right \}_{ij}  \right) \! V_j[c] , \\
	D \! \left( \vec{r} \right) & = D_0 + \frac{v_0^2}{3} \sum_{i,j} \left \{\mathcal{M}^{-1}\right \}_{ij} \!  V_j[c] . 
\end{align}
\end{subequations}
Note, however, that the definition of the matrix~$\mathcal{M}$ is slightly different in two and three dimensions as the rotational noise amplitude~$D_r$ is proportional to the factor~$d-1$, where~$d$ is the actual spatial dimension.

%%%%%%%%%%%%%%%%%%%%%%%%%%%%%%%%%%%%%%%%%%%%%%%%%%%%%%%%%%%%%%%%%%%%%%%%%%%%%%%%%%%%%%
% acknowledgments and references

\begin{acknowledgments}

L.G.N., R.G., and F.P.~acknowledge financial support from Agence Nationale de la Recherche via Grant No.~ANR-15-CE30-0002-01. L.G.N.~was additionally supported by CONACYT PhD scholarship 383881 and R.G. by the People Programme~(Marie Curie Actions) of the European Union's Seventh Framework Programme~(FP7/2007-2013) under REA grant agreement n.~PCOFUND-GA-2013-609102, through the PRESTIGE programme coordinated by Campus France.

\end{acknowledgments}

\begin{center}
	\textbf{AUTHOR CONTRIBUTIONS}
\end{center}

\noindent
L.G.N.~and~R.G.~contributed equally to this work.

% bib
%\bibliography{SPP_ChemoClocks}

\end{document}